\documentclass[english,reprint, citeautoscript, aps, prb, superscriptaddress]{revtex4-1}
\usepackage[T1]{fontenc}
\usepackage[utf8]{inputenc}
\usepackage{color}
\usepackage{verbatim}
\usepackage{algorithm2e}
\usepackage{amsmath}
\usepackage{amssymb}
\usepackage{babel}
\usepackage{graphicx}
\usepackage{filecontents}

\begin{document}
\title{Geometric integration of classical spin dynamics via a mean-field
Schr\"{o}dinger equation}
\author{David~Dahlbom}
\affiliation{Department of Physics and Astronomy, The University of Tennessee,
Knoxville, Tennessee 37996, USA}
\author{Hao~Zhang}
\affiliation{Department of Physics and Astronomy, The University of Tennessee,
Knoxville, Tennessee 37996, USA}
\affiliation{Materials Science and Technology Division, Oak Ridge National Laboratory, Oak Ridge, Tennessee 37831, USA}
\author{Cole Miles}
\affiliation{Department of Physics, Cornell University, Ithaca, New York 14850, USA}
\author{Xiaojian Bai}
\affiliation{
Neutron Scattering Division, Oak Ridge National Laboratory, Oak Ridge, TN 37831, USA}
\author{Cristian~D.~Batista}
\affiliation{Department of Physics and Astronomy, The University of Tennessee,
Knoxville, Tennessee 37996, USA}
\affiliation{Quantum Condensed Matter Division and Shull-Wollan Center, Oak Ridge
National Laboratory, Oak Ridge, Tennessee 37831, USA}
\author{Kipton Barros}
\email{kbarros@lanl.gov}
\affiliation{Theoretical Division and CNLS, Los Alamos National Laboratory, Los
Alamos, New Mexico 87545, USA}

\begin{abstract}
The Landau-Lifshitz equation describes the time-evolution of magnetic
dipoles, and can be derived by taking the classical limit of a quantum
mechanical spin Hamiltonian. To take this limit, one constrains the many-body quantum state to a tensor product of coherent states, thereby neglecting entanglement between sites. Expectation values
of the quantum spin operators
produce the usual classical spin dipoles. One may also consider expectation
values of polynomials of the spin operators, leading to quadrupole
and higher-order spin moments, which satisfy a dynamical equation of motion that generalizes the
Landau-Lifshitz dynamics {[}Zhang and Batista, Phys. Rev. B \textbf{104},
104409 (2021){]}. Here, we reformulate the dynamics of these $N^2-1$ generalized spin components as a mean-field Schr\"odinger equation on the $N$-dimensional coherent state. This viewpoint suggests efficient integration methods that respect the local symplectic structure of the classical spin dynamics.
\end{abstract}
\maketitle
\global\long\def\tr{\mathrm{\mathrm{tr}}\,}%
\global\long\def\mathd{\mathrm{d}}%
\global\long\def\im{\mathrm{i}\,}%
\global\long\def\cay{\mathrm{cay}}%

\section{Introduction}

The Landau-Lifshitz dynamics (LLD),

\begin{equation}
\frac{\mathd\mathbf{s}_{i}}{\mathd t}=-\mathbf{s}_{i}\times\frac{\partial H}{\partial\mathbf{s}_{i}},\label{eq:LL_dynamics}
\end{equation}
describes the time evolution of classical spins $\mathbf{s}_{i}$
with conserved Hamiltonian $H(\mathbf{s}_{1},\dots\mathbf{s}_{L})$. In
the special case of a time-invariant effective field $\mathbf{B}_{i}=-\partial H/\partial\mathbf{s}_{i}$,
each spin $\mathbf s_i$ would simply precess around $\mathbf B_i$.

LLD is one possible classical limit of a quantum mechanical
spin system. Effectively, each spin operator $\hat{S}_{i}^{\alpha}$
is replaced by its expectation value, $s_{i}^{\alpha}=\langle Z|\hat{S}_{i}^{\alpha}|Z\rangle$,
representing the spin angular momentum on site $i$, measured in units
of $\hbar=1$. The quantum state is approximated as a tensor product, $|Z\rangle=|Z_{1}\rangle\otimes\dots\otimes|Z_{L}\rangle$,
thereby neglecting entanglement between sites. Spin operators $\hat{S}_{i}^{\alpha}$
for distinct sites $i\neq j$ commute. In each local Hilbert space
$i$, spin operators act as generators for the Lie group SU(2), which satisfy the commutation relation
\begin{equation}
\left[\hat{S}_{i}^{\alpha},\hat{S}_{i}^{\beta}\right]=\im
\epsilon_{\alpha\beta\gamma}\hat{S}_{i}^{\gamma},
\label{eq:su2_commut}
\end{equation}
where we use the convention of summation over repeated Greek indices (here, $\gamma = 1,2,3$).
The fully antisymmetric Levi-Civita symbol $\epsilon_{\alpha\beta\gamma}$
appearing in this commutator is the underlying source of the vector
cross product appearing in the LLD. An explicit construction
of the spin operators $\hat{S}_{i}^{\alpha}$, in an arbitrary spin-$S$ representation, is presented in Appendix~\ref{sec:spin_rep}.

Our interest is a generalized spin dynamics (GSD) that better approximates local quantum spin states~\cite{Zhang21}. Unlike LLD, which describes only the expected spin dipole $\mathbf s_i$, GSD describes the evolution of a full set of quantum expectation values $\mathbf n_i$ for each local Hilbert space $i$.
% is the study of ``weakly-entangled'' magnetic materials via a  classical approximation that generalizes LLD. 
This generalization is strictly necessary to model large classes of magnets  
with effective spins $S > 1/2$ and strong single-ion
anisotropy induced by the combination of spin-orbit coupling and crystal
field effects, such as  $4d$--$5d$ and $4f$--$5f$-electron materials as well as 
several $3d$ magnets~\cite{Zapf06,Do2020,Bai21}.
The generalization is also necessary to describe magnets comprising weakly-coupled entangled units, such as dimers~\cite{Jaime04}, trimers~\cite{Qiu05} and tetrahedra~\cite{Okamoto13}. Both LLD and GSD are classical approximations that neglect entanglement between different local Hilbert spaces.

Let us now define the generalized spin vector $\mathbf n_i$. Local quantum states $|Z_{i}\rangle$ have dimension $N = 2S+1$ for spins of magnitude $S$. Such states evolve under special unitary transformations, i.e., the Lie group SU($N$). This group is generated by the traceless Hermitian operators, i.e., the Lie algebra $\mathfrak{su}(N)$, with dimension $N^2-1$. An operator basis $\hat T^\alpha_i$ for $\mathfrak{su}(N)$ spans all local physical observables for site $i$. Given an underlying quantum state $|Z_i\rangle$, we define generalized spin components to be the expectation values, $n_{i}^{\alpha}=\langle Z_i|\hat{T}_{i}^{\alpha}|Z_i\rangle$. Without loss of generality, we take $\hat{S}_{i}^{\alpha}$ to be a subset of $\hat{T}_{i}^{\alpha}$, such that the three components
of the spin dipole $s_{i}^{\alpha}$ are a subset of the $N^2-1$ components of $n_i^\alpha$.

% Lie algebra generators are closed under commutation,
% which defines the structure constants $f_{\alpha\beta\gamma}$.
%We define the generalized spin components as expectation values, $n_{i}^{\alpha}=\langle Z_i|\hat{T}_{i}^{\alpha}|Z_i\rangle$.

Generalized spins evolve according to the GSD~\citep{Zhang21},
\begin{equation}
\frac{\mathd n_{i}^{\alpha}}{\mathd t}= f_{\alpha\beta\gamma}\frac{\partial H}{\partial n_{i}^{\beta}}n_{i}^{\gamma},\label{eq:gen_dynamics}
\end{equation}
where the classical Hamiltonian $H(\mathbf n_1, \dots \mathbf n_L)$ is given by the quantum Hamiltonian $\hat{\mathcal H}$ under the substitution rule $\hat T_i^\alpha \rightarrow n_i^\alpha$ (cf. Appendix~\ref{sec:qm_review}). The structure constants $f_{\alpha \beta \gamma}$ are defined by the commutation relation for generators,
\begin{equation}
\left[\hat{T}_{i}^{\alpha},\hat{T}_{i}^{\beta}\right]=\im f_{\alpha\beta\gamma}\hat{T}_{i}^{\gamma}.\label{eq:bracket_op}
\end{equation}
In the special case of $N=2$, one finds $f_{\alpha\beta\gamma}=\epsilon_{\alpha\beta\gamma}$, thereby recovering the LLD of Eq.~\eqref{eq:LL_dynamics}.

This paper is concerned with the efficient numerical integration of  Eq.~(\ref{eq:gen_dynamics}), in a way that respects the underlying geometric structure. We will reformulate GSD as a mean-field Schr\"{o}dinger equation,
\begin{equation}
\frac{\mathd}{\mathd t}\mathbf{Z}_{i}=-\im\mathfrak{H}_{i}\mathbf{Z}_{i},\label{eq:Schrodinger}
\end{equation}
where $\mathbf{Z}_{i}\in\mathbb{C}^{N}$
represents components of $|Z_{i}\rangle$ in some basis, and
\begin{equation}
\mathfrak{\mathfrak{H}}_{i}=\frac{\partial H}{\partial n_{i}^{\alpha}}T^{\alpha}\label{eq:frak_H}
\end{equation}
may be interpreted as an effective local Hamiltonian matrix that acts on $\mathbf Z_i$. Without loss of generality, we have assumed a basis for each local Hilbert space such that the matrix representations $T^{\alpha}$ of the quantum operators $\hat{T}_{i}^{\alpha}$ are independent of site $i$.
The generalized spin components are then
\begin{equation}
n_{i}^{\alpha}=\mathbf{Z}_{i}^{\dagger}T^{\alpha}\mathbf{Z}_{i}.\label{eq:n_basis}
\end{equation}
The derivation of Eq.~(\ref{eq:Schrodinger}) will be presented in
Sec.~\ref{sec:Schrodinger}.

There are great practical advantages to reformulating the dynamics of expectation values $\mathbf n_i(t)$ as a dynamics of underlying coherent states $\mathbf Z_i(t)$. When $N$ is large, it is much preferred to work with the $2N$ real components of the complex vector $\mathbf Z_i$ rather than the $N^2-1$ real components of $\mathbf n_i$. We emphasize that both objects carry the same physical information. The state $\mathbf Z_i$ maps to local physical observables $\mathbf n_i$ via Eq.~\eqref{eq:n_basis}. Conversely, $\mathbf n_i$ is physically valid if and only if there exists a corresponding $\mathbf Z_i$ that satisfies Eq.~\eqref{eq:n_basis}.
Note that an overall complex phase factor, $\mathbf Z_i \rightarrow \mathbf e^{-i \phi} \mathbf Z_i$, is irrelevant to observables $\mathbf n_i$. Also note that the magnitude $|\mathbf Z_i|$ is invariant under the unitary evolution of Eq.~(\ref{eq:Schrodinger}). The remaining physical states $\mathbf Z_i$ live on a $2(N-1)$-dimensional manifold, known mathematically as the complex projective space CP$^{(N-1)}$. The corresponding space of physically allowed spins $\mathbf n_i$, embedded within $\mathbb R^{N^2-1}$, is therefore highly constrained. A simple way to enforce these constraints is to work with  $\mathbf Z_i$ directly, as in Eq.~(\ref{eq:Schrodinger}).

As a concrete example, consider the case of a single spin-1/2 site, with $N=2$ quantum levels (spin up and spin down), and a physical manifold of dimension $2(N-1)=2$. This manifold can be understood through the Bloch sphere construction, which yields normalized dipoles. The Landau-Lifshitz equation defines a dynamics directly on these dipoles, which inherently conserves dipole magnitude. The Schr\"odinger dynamics, Eq.~(\ref{eq:Schrodinger}), defines an equivalent dynamics via the evolution of the quantum state $\mathbf Z \in {\mathbb C}^2$, with conserved magnitude, up to an irrelevant complex phase factor, yielding again two real degrees of freedom. In the mathematics literature, the equivalence between these two dynamics is known as the momentum map~\cite{Marsden99,McLachlan15}.

Equation~(\ref{eq:gen_dynamics}), for general $N$, can be understood as a special type of Lie-Poisson system~\cite{Marsden99}. The geometric meaning of a Lie-Poisson system (for antisymmetric structure constants  $f_{\alpha \beta \gamma}$) is perhaps best understood through the matrix $\mathfrak{n}_{i}= n_{i}^{\alpha}T^{\alpha}$, known in high-energy physics as the color field. As we will review in Sec.~\ref{sec:Schrodinger}, this matrix evolves dynamically as $\mathd \mathfrak{n}_{i} /\mathd t = \im [ \mathfrak{n}_{i}, \mathfrak{H}_{i}]$, and its eigenvalues are constants of motion. Numerical methods have recently been designed to exactly respect this isospectral flow~\cite{Modin20,Viviani20}. In our specific context, there is a unique greatest eigenvalue of $\mathfrak{n}_{i}$, and the associated eigenvector is $\mathbf Z_i$, up to an irrelevant scaling factor. All other eigenvalues of $\mathfrak{n}_i$ are degenerate. The isospectral flow condition becomes equivalent to the constraints implicit in Eq.~(\ref{eq:n_basis}). Our approach therefore reformulates the full matrix dynamics of $\mathfrak{n}_i$ as the dynamics of the single eigenvector $\mathbf Z_i$. The final scheme, which we call the Schr\"odinger midpoint method, will be presented in Sec.~\ref{sec:numerics}.

An important property of Eq.~(\ref{eq:Schrodinger}) is that it has a canonical Hamiltonian structure for any $N$. Specifically, the real and imaginary components of $\mathbf{Z}_{i}$ act as canonical positions and momenta, and obey Hamilton's equations of motion. The Schr\"odinger midpoint method exactly respects this symplectic structure and therefore enables dynamical integration over arbitrarily long time-scales without numerical drift.

Given a specific quantum Hamiltonian $\hat{\mathcal{H}}$, the numerical implementation of the Schr\"odinger midpoint method is relatively straightforward. The main task is to build the matrix $\mathfrak H_i(\mathbf n_1, \dots \mathbf n_L)$ for each site $i$. Using the framework of Appendix~\ref{sec:qm_review}, the classical Hamiltonian $H(\mathbf n_1, \dots \mathbf n_L)$ will be at most linear in each spin component $n_i^\alpha$. Then $\mathfrak H_i$ can be viewed as a mean-field approximation to the quantum Hamiltonian $\hat{\mathcal H}$ under the substitution $\hat{T}_j^\alpha \rightarrow n_j^\alpha$ for all sites $j \neq i$, up to an irrelevant constant shift. The linear combination of generators $T^\alpha$ appearing in Eq.~\eqref{eq:frak_H} becomes a simple polynomial of spin operators, directly reflecting the definition of $\hat{\mathcal H}$. We will demonstrate this procedure through explicit examples in Sec.~\ref{sec:benchmarks}.

\section{Classical dynamics in the Schr\"{o}dinger picture\label{sec:Schrodinger}}

\subsection{Unitary evolution of expectation values\label{sec:unitary}}

Equation~(\ref{eq:gen_dynamics}) is a Lie-Poisson system, and describes
co-adjoint orbits on the dual Lie algebra~\citep{Marsden99,Engo01}.
We will make this statement concrete using ordinary matrix language.

Let $T^{\alpha}$ be generators for SU($N$) in the defining representation. That is, $T^\alpha$ are a basis for the Lie algebra $\mathfrak{su}(N)$, the space of traceless, Hermitian, $N\times N$ matrices. The matrix
commutator is, 
\begin{equation}
[T^{\alpha},T^{\beta}]=\im f_{\alpha\beta\gamma}T^{\gamma},\label{eq:bracket}
\end{equation}
inherited from that of the quantum operators, Eq.~(\ref{eq:bracket_op}).

We will require that the basis satisfies an orthonormality condition,

\begin{equation}
\mathrm{tr}\,T^{\alpha}T^{\beta}=\tau\delta_{\alpha\beta}.\label{eq:Lie_ortho}
\end{equation}
This condition makes it possible to interpret $T^{\alpha}$ as a basis
also for the dual Lie algebra, $\mathfrak{su}^{\ast}(N)$. Orthonormality
is equivalent to antisymmetry of $f_{\alpha\beta\gamma}$ in all indices
\begin{equation}
f_{\alpha\beta\gamma}=-f_{\beta\alpha\gamma}=-f_{\alpha\gamma\beta}.\label{eq:f_anti}
\end{equation}

Our convention is to select a basis $T^{\alpha}$ that includes the three spin matrices $S^{\alpha}$
as a subset. Substitution of $T^{\alpha}\rightarrow S^{\alpha}$ into
Eq.~(\ref{eq:Lie_ortho}) determines $\tau$, as given in Eq.~(\ref{eq:tau_def})
of Appendix~\ref{sec:spin_rep}.

Equation~(\ref{eq:gen_dynamics}) defines the dynamics of spin components
$n_{i}^{\alpha}$. Equivalently, we may consider the time evolution
of the matrix
\begin{equation}
\mathfrak{n}_{i}= n_{i}^{\alpha}T^{\alpha},\label{eq:frak_n}
\end{equation}
interpreted as an element of dual Lie algebra $\mathfrak{su}^{\ast}(N)$.

It will be convenient to identify the energy gradient $\partial H/\partial n_{i}^{\alpha}$
with the matrix $\mathfrak{\mathfrak{H}}_{i}$ defined in Eq.~(\ref{eq:frak_H}),
interpreted as an element of $\mathfrak{su}(N)$. Using this notation,
Eq.~(\ref{eq:gen_dynamics}) is compactly expressed as
\begin{equation}
\frac{\mathd\mathfrak{n}_{i}}{\mathd t}=\im\left[\mathfrak{n}_{i},\mathfrak{H}_{i}\right],\label{eq:frak_n_dyn}
\end{equation}
which follows from the antisymmetry of $f_{\alpha\beta\gamma}$, and
the matrix commutators in Eq.~(\ref{eq:bracket}). The dynamics
may also be written
\begin{equation}
\mathfrak{n}_{i}(t)=U_{i}(t)\mathfrak{n}_{i}(0)U_{i}^{-1}(t),\label{eq:frak_n_t}
\end{equation}
where $U(t)\in\mathrm{SU}(N)$ satisfies
\begin{equation}
\frac{\mathd}{\mathd t}U_{i}(t)=-\im\mathfrak{H}_{i}(t)U_{i}(t),\label{eq:dU_dt}
\end{equation}
with initial condition $U(0)=I$, as may be verified by explicit
differentiation.%
\begin{comment}
\begin{align*}
\dot{\mathfrak{n}} & =\dot{U}\mathfrak{n}_{0}U^{-1}-U\mathfrak{n}_{0}U^{-1}\dot{U}U^{-1}\\
 & =-\im\left(\mathfrak{H}U\mathfrak{n}_{0}U^{-1}-U\mathfrak{n}_{0}U^{-1}\mathfrak{H}\right)\\
 & =-\im\left(\mathfrak{H}\mathfrak{n}-\mathfrak{n}\mathfrak{H}\right)\\
 & =+\im\left[\mathfrak{n},\mathfrak{H}\right].
\end{align*}
\end{comment}

\begin{comment}
This unitary evolution of $\mathfrak{n}_{i}(t)$ is a concrete realization
of a more general mathematical statement: Lie-Poisson systems such
as Eq.~(\ref{eq:gen_dynamics}) describe the coadjoint orbit of a
dual Lie algebra element under the group action.
\end{comment}

\subsection{Schr\"{o}dinger dynamics of coherent states\label{sec:coherent_states}}

Now we will reformulate the unitary evolution of the matrix
$\mathfrak{n}_{i}$ as a dynamics of the vector $\mathbf{Z}_{i}$ that gives rise to expectation values $n_i$ via Eq.~(\ref{eq:n_basis}).
% The key observation is that $\mathbf{Z}_{i}$ an eigenvector of $\mathfrak{n}_{i}$.
To derive this dynamics in a way that
builds physical intuition, we introduce the outer product,
\begin{equation}
\rho_{i}=\mathbf{Z}_{i}\mathbf{Z}_{i}^{\dagger},\label{eq:rho_def}
\end{equation}
in analogy with a pure density matrix for site $i$.

Spin components, Eq.~\eqref{eq:n_basis},
may be calculated in two different ways,
\begin{equation}
\mathrm{tr}\,\rho_{i}T^{\alpha}=n_{i}^{\alpha}=\mathrm{tr}\,\mathfrak{n}_{i}T^{\alpha}/\tau.\label{eq:tr_rho_T}
\end{equation}
For the second equality, we used the definition of Eq.~(\ref{eq:frak_n})
and orthonormality, Eq.~(\ref{eq:Lie_ortho}).

The matrix $\rho_{i}$ is Hermitian, and the generators $T^{\alpha}$
span all $N\times N$ Hermitian, traceless matrices. From this we deduce
\begin{equation}
\rho_{i}=\mathfrak{n}_{i}/\tau+ cI,\label{eq:rho_to_frak_n}
\end{equation}
where $I$ is the identity matrix, and $c = |\mathbf Z_i|^2/N$.

Because $I$ commutes with any matrix, the matrices $\mathfrak{n}_{i}$
and $\rho_{i}$ share the same dynamical equation
\begin{equation}
\frac{\mathd\rho_{i}}{\mathd t}=\im\left[\rho_{i},\mathfrak{H}_{i}\right],\label{eq:drho_dt}
\end{equation}
or equivalently,
\begin{equation}
\mathfrak{\rho}_{i}(t)=U_{i}(t)\rho_{i}(0)U_{i}^{-1}(t).\label{eq:rho_unitary}
\end{equation}
This dynamics may be interpreted as the von Neumann evolution of the
density matrix.
% [note the the missing minus sign relative to the Heisenberg equation of motion, cf. Eq.~(\ref{eq:dA_dt}) in Appendix~\ref{sec:qm_review}].
Referring to Eq.~(\ref{eq:rho_def}), it follows that the coherent states must evolve as,
\begin{equation}
\mathbf{Z}_{i}(t)=U_{i}(t)\mathbf{Z}_{i}(0).
\end{equation}
Taking the time derivative of both sides, and substituting from Eq.~(\ref{eq:dU_dt})
yields 
\[
\mathd\mathbf{Z}_{i}/\mathd t=-\im\mathfrak{H}_{i}\mathbf{Z}_{i},
\]
which confirms our claim that Eq.~(\ref{eq:Schrodinger}) is a reformulation of the GSD defined in Eq.~(\ref{eq:gen_dynamics}).

% Extending the analogy with quantum mechanics, we will refer to $\mathfrak{H}_{i}(t)$
% as the \emph{local Hamiltonian} on site $i$. In contrast, recall that $\mathfrak{H}_{i}(t)$
% was defined in Eq.~(\ref{eq:frak_H}) as the embedding of the classical
% Hamiltonian gradient $\partial H/\partial n_{i}^{\alpha}$ into the Lie
% algebra $\mathfrak{su}(N)$. To understand how these two viewpoints
% are compatible, we refer to Appendix~\ref{sec:qm_review}. For the generalized spin dynamics in the fundamental representation of SU($N$), the classical Hamiltonian $H$ of Eq.~(\ref{eq:E_def})
% is obtained from the many-body quantum Hamiltonian $\hat{\mathcal{H}}$
% under the substitution $\hat{T}_{i}^{\alpha}\rightarrow n_{i}^{\alpha}$.
% Furthermore, the spin components $n_{i}^{\alpha}$ for each site $i$
% appear at most linearly in each term of $H$. Then $\mathfrak{H}_{i}$
% may be viewed as a semiclassical\emph{ }approximation to $\hat{\mathcal{H}}$,
% in which we make the substitution $\hat{T}_{j}^{\alpha}\rightarrow n_{j}^{\alpha}$
% only for sites $j\neq i$.

The results of this section may be restated in a more abstract and
general mathematical language. Equation~(\ref{eq:frak_n_dyn}) can
be viewed as an isospectral flow $\mathd W/\mathd t=[B,W]$, with $W=\mathfrak{n}_{i}$
and $B=(\im\mathfrak{H}_{i})^{\dagger}$.
The eigenvalues of $W$ are constants of motion. Each eigenvector $\mathbf{v}$
of $W$ satisfies the dynamical equation $\mathd\mathbf{v}/\mathd t=B\mathbf{v}$,
corresponding to our Schr\"{o}dinger equation. Although we derived this
result in the context of the Lie-Poisson system on $\mathfrak{su}(N)$,
it generalizes to a much broader class of so-called \emph{reductive
}Lie algebras~\citep{Modin20,Viviani20}. Note that most classical
Lie algebras are reductive, including $\mathfrak{\mathfrak{gl}}(N,\mathbb{C})$,
$\mathfrak{gl}(N,\mathbb{R})$, $\mathfrak{so}(N)$, and $\mathfrak{sp}(N)$,
in addition to our working example of $\mathfrak{su}(N)$. In our
application to spin dynamics, we benefit from the fact that a single
eigenvector, $\mathbf{v}=\mathbf{Z}_{i}$, fully describes the matrix
$W=\mathfrak{n}_{i}$. More generally, if the initial condition $W(0)$
has low rank (up to some constant shift), then
the time-evolved state $W(t)$ will continue to have low rank, and
modeling the dynamics through the time-evolving eigenvectors becomes
beneficial.

\subsection{Conservation laws \label{sec:conservation_laws}}

Lie-Poisson systems such as Eq.~\eqref{eq:gen_dynamics} satisfy a number of conservation laws. Some of these are associated with the geometric structure of the phase space, and are independent of the choice of Hamiltonian. One may verify that any function $C(\mathbf{n}_{i})$ that satisfies
$(\partial C/\partial n_{i}^{\alpha})f_{\alpha\beta\gamma}n_{i}^{\gamma}=0$
is a constant of motion. Such functions arise from Casimirs of the Lie algebra, i.e., symmetric, homogeneous polynomials of the basis
matrices $T^{\alpha}$ that commute with all algebra elements. There are $N-1$ Casimirs of SU($N$). The simplest example is the quadratic Casimir $|\mathbf n_i|^2 = \sum_{\alpha}(n_{i}^{\alpha})^{2}$
which, for spin-$\frac{1}{2}$ systems, reduces to the dipole magnitude squared.

A second class of conservation laws arise from symmetries of the Hamiltonian. For example, energy is a constant of motion provided that the Hamiltonian has no explicit time-dependence. One may verify $\mathd H/\mathd t=0$ directly by contracting
$\partial H/\partial n_{i}^{\alpha}$ on both sides of Eq.~(\ref{eq:gen_dynamics}),
and using the antisymmetry of $f_{\alpha\beta\gamma}$.

The Schr\"{o}dinger dynamics, Eq.~(\ref{eq:Schrodinger}), is equivalent
to the generalized spin dynamics, Eq.~(\ref{eq:gen_dynamics}),
and therefore shares these conservation laws.

Equation~\eqref{eq:gen_dynamics} has the form of a Lie-Poisson system.
The Darboux-Lie theorem states that Lie-Poisson systems have a local
Hamiltonian structure~\cite{Marsden99,Hairer06}.
That is, in the neighborhood of
each spin configuration $n_{i}^{\alpha}$, there exists a change of
coordinates that gives rise to canonical variables $(\mathbf{p}_{i},\mathbf{q}_{i})$
that satisfy Hamilton's equations of motion locally. Interestingly, the Schr\"{o}dinger equation gives
rise to a Hamiltonian dynamics that is valid \emph{globally}. Specifically,
in Appendix~\ref{sec:schro_canonical} we demonstrate that the real
and imaginary components of the coherent state, 
\begin{equation}
\mathbf{Z}_{i}=(\mathbf{p}_{i}-\im\mathbf{q}_i)/\sqrt{2},\label{eq:Z_canonical}
\end{equation}
satisfy Hamilton's equations of motion,
\begin{equation}
\frac{\mathd\mathbf{p}_{i}}{\mathd t}=-\frac{\partial H}{\partial\mathbf{q}_{i}},\quad\frac{\mathd\mathbf{q}_{i}}{\mathd t}=+\frac{\partial H}{\partial\mathbf{p}_{i}}.\label{eq:hamiltons_equations}
\end{equation}
The canonical Hamiltonian structure of the Schr\"{o}dinger equation
ensures conservation of the symplectic 2-form $\sum_{i,a}\mathd p_{i,a}\wedge\mathd q_{i,a}$.

\subsection{State normalization\label{sec:coherent_state_normalization}}

The magnitude of $\mathbf Z_i$ is a conserved quantity in the Schr\"odinger dynamics, Eq.~\eqref{eq:Schrodinger}. Absent other knowledge, the normalization convention $|\mathbf Z_i| = 1$ is a natural choice, and emphasizes the interpretation of $\mathbf Z_i$ as a quantum mechanical coherent spin state. Rescaling $\mathbf Z_i$ can be useful, however, to adjust the overall magnitude of the classical spin components. A carefully selected rescaling can strongly enhance the agreement between an approximate classical model and the true quantum mechanical system~\cite{Huberman08}. 

Consider, first, the LLD of spin dipoles, Eq.~\eqref{eq:LL_dynamics}. This coincides with Eq.~\eqref{eq:gen_dynamics} on the Lie algebra $\mathfrak{su}(2)$, such that the matrices $T^\alpha$ appearing in Eq.~\eqref{eq:frak_H} are the three generators $S^\alpha$ of SU(2). Although we have so far interpreted $T^\alpha$ as generators in the fundamental representation of SU($N$), this assumption is unnecessary.
In particular, we can faithfully describe the LLD using generators $T^\alpha$ in any irreducible representation of SU(2), labeled by spin $S \in \{\frac{1}{2}, 1, \frac{3}{2} \dots \}$. Each dipole magnitude $|\mathbf s_i|$ is a conserved quantity under LLD, and takes the value $S |\mathbf Z_i|^2$ in the spin-$S$ representation. To model a different dipole magnitude $|\mathbf s_i| = s_0$, we should normalize $\mathbf Z_i$ such that
\begin{equation}
|\mathbf Z_i|^2 = s_0 / S \label{eq:Z_normalization}
\end{equation}

Consider, second, the GSD, Eq.~\eqref{eq:gen_dynamics}, interpreted as the evolution of $N^2-1$ spin components $n_i^\alpha$ in the fundamental representation of SU($N$). Equivalently, the Schr\"odinger picture describes this dynamics as the unitary evolution of coherent spin states $\mathbf Z_i$. Our convention to take the spin matrices $S^\alpha$ as a subset of the orthonormal generators $T^\alpha$ imposes the normalization $|\mathbf n_i| = S |\mathbf Z_i|^2$, where $S = (N-1)/2$. One may wish to select the normalization of $\mathbf Z_i$ according to a quantum mechanical sum rule derived from the SU($N$) quadratic Casimir.

\section{Numerical Methods\label{sec:numerics}}

Geometric integration\emph{ }aims to approximate the flow of a dynamical
system while exactly satisfying geometrical constraints~\citep{Iserles00,Iserles16,Hairer06}.
For example, Eq.~(\ref{eq:rho_unitary})
suggests that one integration time-step should take the form~\citep{Engo01}
\begin{equation}
\rho_{i}(t)\rightarrow\rho_{i}(t+\Delta t)=U_{t}\rho_{i}(t)U_{t}^{-1}.\label{eq:rho_unitary_disc}
\end{equation}
To obtain an exactly unitary matrix $U_{t}$ that approximates the integral of Eq.~(\ref{eq:dU_dt}),
one may use, e.g., the method of Runge-Kutte Munthe-Kaas~\citep{Munthe-Kaas98}.
This unitary evolution ensures conservation of Casimirs, but
conservation of other geometrical properties, such as the local symplectic
2-form, is not automatically guaranteed.

Designing efficient symplectic integration schemes for Lie-Poisson
systems is a topic of considerable interest~\citep{Zhong88,Channell91,McLachlan93,McLachlan95}.
A common strategy is to employ operator splitting and the fact that
the composition of symplectic maps is again symplectic. For example,
in the context of LLD, one may partition the spins into
non-interacting groups, and cycle through symplectic updates
for each of these groups using symmetric Strang splitting (i.e., a
Suzuki-Trotter type decomposition)~\citep{Krech98,Omelyan01,Tranchida18}.

Alternatively, one may seek symplectic integrators for Lie-Poisson
systems that update all dynamical variables simultaneously, and in
a symmetric way. The spherical midpoint method, designed specifically for LLD of dipoles, is one example~\citep{McLachlan14}. Quite recently, Modin
and Viviani introduced the class of Isospectral Symplectic Runge--Kutta methods (IsoSyRK)~\citep{Modin20}, which applies to any Lie-Poisson
system on a reductive Lie algebra; this covers most Lie-Poisson systems
of practical interest, including generalized spin dynamics on $\mathfrak{su}(N)$,
and the LLD as a special case. The Isospectral Minimal Midpoint (IMM) method is a particularly elegant variant of IsoSyRK~\citep{Viviani20}.

The standard implicit midpoint
method is known to be symplectic when applied to canonical Hamiltonian systems~\cite{Hairer06}. We will next demonstrate that the implicit midpoint method applied to the Schr\"{o}dinger equation~(\ref{eq:Schrodinger}) coincides with IMM method applied to the equivalent Lie-Poisson system, Eq.~(\ref{eq:gen_dynamics}).

\subsection{Schr\"{o}dinger midpoint}\label{subsec:Schrodinger_midpoint}

To integrate Eq.~(\ref{eq:Schrodinger}) over one time-step $\mathbf{Z}_{i}\rightarrow\mathbf{Z}_{i}'$,
the implicit midpoint method has the symmetric form
\begin{equation}
\frac{\mathbf{Z}_{i}'-\mathbf{Z}_{i}}{\Delta t}=-\im\mathfrak{\tilde{H}}_{i}\tilde{\mathbf{Z}}_{i},\label{eq:midpoint}
\end{equation}
The right-hand side involves the midpoint state,
\begin{equation}
\tilde{\mathbf{Z}}_{i}=\frac{\mathbf{Z}_{i}'+\mathbf{Z}_{i}}{2}.\label{eq:Zbar}
\end{equation}
The symbol $\mathfrak{\tilde{H}}_{i}$ denotes the local Hamiltonian
of Eq.~(\ref{eq:frak_H}), evaluated as a function of $\tilde{\mathbf{Z}}_{j}$
at all sites $j$.

The self-consistent value of $\mathbf{Z}_{i}'$ may be calculated
numerically as follows. Starting with an initial guess $\mathbf{Z}_{i;0}'=\mathbf{Z}_{i}$,
we iteratively calculate
\begin{align}
\tilde{\mathbf{Z}}_{i;k} & =\frac{1}{2}\left(\mathbf{Z}'_{i;k}+\mathbf{Z}_{i}\right),\label{eq:zbar_update}\\
\mathbf{Z}'_{i;k+1} & =\mathbf{Z}_{i}-\im\Delta t\mathfrak{\tilde{H}}_{i;k}\tilde{\mathbf{Z}}_{i;k},\label{eq:zprime_update}
\end{align}
with $\mathfrak{\tilde{H}}_{i;k}$ defined in the natural way. Iterations
terminate when $\mathbf{Z}'_{i;k}$ has converged within numerical
tolerance. For example, at 64-bit floating point precision, we may
require $|\mathbf{Z}_{i;k+1}^{'}-\mathbf{Z}_{i;k}^{'}|<10^{-14}$.
This condition is typically satisfied in $k\lesssim10$ iterations,
given a reasonably small step size $\Delta t$.

The Schr\"{o}dinger equation is a canonical Hamiltonian
system, via Eqs.~(\ref{eq:Z_canonical}) and~(\ref{eq:hamiltons_equations}).
For such systems, the implicit midpoint method is known to be a symplectic integrator~\citep{Hairer06}.

The Schr\"{o}dinger midpoint method is norm preserving. To see this, we
left-multiply both sides of Eq.~(\ref{eq:midpoint}) by $\tilde{\mathbf{Z}}_{i}^{\dagger}$,
\begin{equation}
\frac{1}{2\Delta t}\left(\mathbf{Z}_{i}'+\mathbf{Z}_{i}\right)^{\dagger}\left(\mathbf{Z}_{i}'-\mathbf{Z}_{i}\right)=-i\tilde{\mathbf{Z}}_{i}^{\dagger}\mathfrak{\tilde{H}}_{i}\tilde{\mathbf{Z}}_{i}.
\end{equation}
The right-hand side is purely imaginary, since $\tilde{\mathfrak{H}}_{i}$
is Hermitian. Setting the real terms on the left-hand side to zero,
we find $|\mathbf{Z}_{i}|^2=|\mathbf{Z}_{i}'|^2$.

We will now demonstrate that the Schr\"{o}dinger midpoint method is an instance of the IMM method~\citep{Modin20,Viviani20}. The density matrix $\rho_i = \mathbf{Z}_i \mathbf{Z}_i^\dagger$ evolves according to Eq.~\eqref{eq:drho_dt}. This dynamics can be understood as an isospectral flow $\mathd W/\mathd t=[B(W), W]$ where $B =-\im\mathfrak{H}_{i}$ and $W = \rho_i$. One time-step $W \rightarrow W'$ of the IMM method is defined as,
\begin{align}
W & =\left(I-\frac{\Delta t}{2}B(\tilde{W})\right)\tilde{W}\left(I+\frac{\Delta t}{2}B(\tilde{W})\right)\label{eq:spectral1}\\
W' & =\left(I+\frac{\Delta t}{2}B(\tilde{W})\right)\tilde{W}\left(I-\frac{\Delta t}{2}B(\tilde{W})\right).\label{eq:spectral2}
\end{align}
where the midpoint state $\tilde W$ and final state $W'$ are to be solved self-consistently.

Equations~\eqref{eq:midpoint} and~\eqref{eq:Zbar} may be rewritten as
\begin{align}
\mathbf{Z}_{i} & =\left(I+\im\frac{\Delta t}{2}\tilde{\mathfrak{H}}_{i}\right)\tilde{\mathbf{Z}}_{i} \label{eq:Z_mid_back} \\
\mathbf{Z}'_{i} & =\left(I-\im\frac{\Delta t}{2}\tilde{\mathfrak{H}}_{i}\right)\tilde{\mathbf{Z}}_{i}. \label{eq:Z_mid_forward}
\end{align}
Intuitively, this says that the midpoint state $\tilde{\mathbf{Z}}_{i}$ can be obtained either by integrating forward from $\mathbf{Z}_{i}$, or backward from $\mathbf{Z}'_{i}$. Calculating the outer products $W=\mathbf{Z}_{i}\mathbf{Z}_{i}^{\dagger}$ and $W'=\mathbf{Z}'_{i}\mathbf{Z}_{i}'^{\dagger}$, we exactly reproduce the IMM equations. Note that Eq.~\eqref{eq:Zbar} defines $\tilde{\mathbf{Z}}_{i}$ as a simple vector average of the initial and final states, whereas $\tilde W = \tilde{\mathbf{Z}}_{i}\tilde{\mathbf{Z}}_{i}^{\dagger}$ cannot be expressed that way.

\subsection{Schr\"{o}dinger midpoint applied to the LLD \label{sec:schr_ll}}

The LLD, Eq.~\eqref{eq:LL_dynamics}, is a special case of the GSD, Eq.~\eqref{eq:gen_dynamics}. It can therefore be formulated as a Schr\"{o}dinger equation on an SU(2) representation, and integrated using the midpoint method, Eqs.~\eqref{eq:midpoint} and~\eqref{eq:Zbar}. In the special cases of spin $S=\frac{1}{2}$ and $S=1$ representations, the Schr\"{o}dinger midpoint method may be equivalently reformulated as an update rule operating directly on spin dipoles, $\mathbf{s}_i \rightarrow \mathbf{s}'_i$. The final result, derived in Appendix~\ref{sec:schr_ll_derivation}, is
\begin{align}
\mathbf{s}_{i} &=\tilde{\mathbf{s}}_{i}+\frac{\Delta t}{2}\tilde{\mathbf{s}}_{i}\times\frac{\partial H}{\partial\tilde{\mathbf{s}}_{i}}-\frac{\Delta t^{2}}{4}f(\tilde{\mathbf{s}}_{i}) \label{eq:sch_ll_1}\\
\mathbf{s}_{i}'&=\tilde{\mathbf{s}}_{i}-\frac{\Delta t}{2}\tilde{\mathbf{s}}_{i}\times\frac{\partial H}{\partial\tilde{\mathbf{s}}_{i}}-\frac{\Delta t^{2}}{4}f(\tilde{\mathbf{s}}_{i}),\label{eq:sch_ll_2}
\end{align}
where the quadratic correction,
\begin{equation}
f(\tilde{\mathbf{s}}_{i})=\begin{cases}
\frac{1}{2}\frac{\partial H}{\partial\tilde{\mathbf{s}}_{i}}\left(\frac{\partial H}{\partial\tilde{\mathbf{s}}_{i}}\cdot\tilde{\mathbf{s}}_{i}\right)-\frac{1}{4}\left|\frac{\partial H}{\partial\tilde{\mathbf{s}}_{i}}\right|^{2}\tilde{\mathbf{s}}_{i} & \textrm{(spin-$\frac{1}{2}$)}\\
\frac{\partial H}{\partial\tilde{\mathbf{s}}_{i}}\left(\frac{\partial H}{\partial\tilde{\mathbf{s}}_{i}}\cdot\tilde{\mathbf{s}}_{i}\right) & \textrm{(spin-1)}
\end{cases},
\end{equation}
depends on the choice of SU(2) representation. The midpoint state $\tilde{\mathbf s}_i$ can be solved self-consistently from the initial state $\mathbf{s}_i$ using Eq.~\eqref{eq:sch_ll_1} alone. We emphasize that $\tilde{\mathbf s}_i$ is \emph{not} a simple average of $\mathbf{s}_i$ and $\mathbf{s}'_i$. Once $\tilde{\mathbf s}_i$ is known, the final state $\mathbf{s}'_i$ can be solved directly using Eq.~\eqref{eq:sch_ll_2}.

The spin-$\frac{1}{2}$ variant of the Schr\"odinger dynamics coincides with the framework presented in Ref.~\onlinecite{McLachlan15}, albeit in a different language. In our notation, the expected dipole is $s_i^\alpha = \mathbf Z_i^\dagger \sigma^\alpha \mathbf Z_i / 2$, where $\sigma^\alpha$ are the Pauli spin matrices. In the math literature, this functional dependence $\mathbf s_i(\mathbf Z_i)$ is known as the \emph{momentum map}~\cite{Marsden99}. The functional dependence $H(\mathbf Z_i)$ is called the \emph{collective Hamiltonian}.

The spin-1 variant of Eqs.~\eqref{eq:sch_ll_1} and~\eqref{eq:sch_ll_2} was previously derived in Ref.~\onlinecite{Viviani20} by applying the IMM method to the Lie-Poisson system on $\mathfrak{so}(3)$. Recall that the generators of SO(3) in its defining representation are also generators of SU(2) in its spin-1 representation.

\subsection{Spherical midpoint}
The spherical midpoint method is a powerful symplectic integrator for LLD~\citep{McLachlan14}. One integration
time-step $\mathbf{s} \rightarrow \mathbf{s}'$ is defined by the update rule
\begin{equation}
\frac{\mathbf{s}_i'-\mathbf{s}_i}{\Delta t}=-\bar{\mathbf{s}}_{i}\times\frac{\partial H}{\partial\bar{\mathbf{s}}_{i}},
\end{equation}
involving the normalized midpoint dipole,
\begin{equation}
\bar{\mathbf{s}}_i=\frac{\mathbf{s}'_i+\mathbf{s}_i}{|\mathbf{s}'_i+\mathbf{s}_i|}.
\end{equation}
The classical Hamiltonian on the right-hand side $H(\bar{\mathbf{s}}_1,\dots \bar{\mathbf{s}}_L)$ is evaluated at the midpoint spin configuration.

The new spin $\mathbf{s}'$ can be solved
self-consistently using iterations analogous to Eqs.~(\ref{eq:zbar_update})
and~(\ref{eq:zprime_update}). Here, however, normalization of the midpoint spin $\bar{\mathbf{s}}$ is employed, which is crucial to the good properties of the method. Various proofs of the symplectic structure have been given, and are somewhat involved~\cite{McLachlan17}.

\subsection{Heun-projected (HeunP)}

All methods above are symplectic. For purposes of benchmarking, we will also compare with the Heun method, a non-symplectic Runge-Kutta
scheme of second order. Applied to Eq.~(\ref{eq:Schrodinger}),
one full integration time-step is
\begin{align}
\mathbf{Z}_{i}^{(1)} & =\mathbf{Z}_{i}-\im\Delta t\mathfrak{H}_{i}\mathbf{Z}_{i}\\
\mathbf{Z}_{i}^{(2)} & =\mathbf{Z}_{i}-\frac{\mathrm{i}\Delta t}{2}\left(\mathfrak{H}_{i}\mathbf{Z}_{i}+\mathfrak{H}_{i}^{(1)}\mathbf{Z}_{i}^{(1)}\right),\\
\mathbf{Z}_{i}' & =\mathbf{Z}_{i}^{(2)}/|\mathbf{Z}_{i}^{(2)}|.
\end{align}
The first step can be interpreted as an explicit Euler predictor
for the update. The second, corrector step involves the local Hamiltonian
$\mathfrak{H}_{i}^{(1)}$ evaluated at $\mathbf{Z}_{j}^{(1)}$ for
all $j$. Finally, the output $\mathbf{Z}_{i}'$ is normalized using
a projection step. We use the name HeunP in reference to prior work
that employed the same scheme to integrate LLD, where the dipoles were the dynamical variables~\citep{Skubic08,Mentink10}.

\section{Numerical benchmarks\label{sec:benchmarks}}

\subsection{Model definitions \label{sec:benchmarks_models}}

For our numerical examples, we consider
the 1D Heisenberg spin chain with an easy-axis anisotropy. We start from the quantum Hamiltonian,
\begin{equation}
\hat{\mathcal{H}}= J \sum_{i=1}^{L}\hat{\mathbf{S}}_{i}\cdot\hat{\mathbf{S}}_{i+1} + D\sum_{i=1}^{L}(\hat{S}_{i}^{z})^{2}, \label{eq:H_spin_chain}
\end{equation}
where $\hat{S}^\alpha$ denote spin operators in the spin-1 representation. We will study this model in two classical limits.

The LLD, Eq.~\eqref{eq:LL_dynamics}, retains only the spin dipole degrees of freedom. Its classical Hamiltonian
\begin{equation}
H^{\mathrm{LLD}}= J \sum_{i=1}^{L}\mathbf{s}_{i}\cdot\mathbf{s}_{i+1} + D\sum_{i=1}^{L}(s_{i}^{z})^{2},
\end{equation}
is obtained from $\hat{\mathcal{H}}$ under the substitution rule $\hat{\mathbf{S}}_{i}\rightarrow \mathbf{s}_{i}$. We will employ the normalization convention $|\mathbf s_i| = 1$. 
The LLD can be formulated as a Schr\"odinger dynamics in some spin-$S$ representation of SU(2). Per Eq.~\eqref{eq:Z_normalization}, the proper normalization of states is $|\mathbf Z_i|^2 = 1/S$.

The GSD of Eq.~\eqref{eq:gen_dynamics} describes an alternative classical limit. This approach allows modeling each spin-1 state in a more physically correct way, and is especially important when there is strong single-ion anisotropy (here, large $D$).  Each spin dipole $\mathbf s_i$ generalizes to an eight component vector $\mathbf n_i$ that includes the original dipole $\mathbf s_i$, as well as five additional quadrupole components. The local coherent state $\mathbf Z_i \in \mathbb{C}^3$ carries information equivalent to the generalized spin $\mathbf n_i \in \mathbb{R}^8$.

Following the arguments of Appendix~\ref{sec:qm_review}, the classical Hamiltonian is obtained from $\hat{\mathcal H}$ under the substitution $\hat T_i^\alpha \rightarrow n_i^\alpha$, with the result,
\begin{equation}
H^{\mathrm{GSD}}= J \sum_{i=1}^{L}\mathbf{s}_{i}\cdot\mathbf{s}_{i+1} + D\sum_{i=1}^{L} c_{\alpha}n_{i}^{\alpha},
\end{equation}
up to an irrelevant constant shift. We have used the generic expansion,
\begin{equation}
(S^{z})^{2} = c_0 I + c_{\alpha}T^{\alpha}.\label{eq:aniso_expand}
\end{equation}
where the spin-1 matrices $S^\alpha$ are a subset of the generators $T^{\alpha}$ of SU(3) in the defining representation. Any single-ion anisotropy term could be expanded in this manner.
The coefficients $c_{\alpha}$
may be calculated explicitly using Eq.~(\ref{eq:Lie_ortho}), but
doing so is not needed for the numerics.

Substitution of $H^{\mathrm{GSD}}$ into Eq.~(\ref{eq:gen_dynamics})
defines the generalized spin dynamics. In practice, we will use the Schr\"{o}dinger formulation, involving the local Hamiltonian of Eq.~\eqref{eq:frak_H},
\begin{align}
\mathfrak{H}^{\mathrm{GSD}}_{i} &= \frac{\partial H^\mathrm{GSD}}{\partial n_{i}^{\alpha}} T^{\alpha} \\
 &= J (s_{i-1}^{\alpha}+s_{i+1}^{\alpha})S^{\alpha} + D c_\alpha T^\alpha.
\end{align}
In a numerical implementation, it is convenient to now undo the expansion of Eq.~(\ref{eq:aniso_expand}),
\begin{equation}
\mathfrak{H}^{\mathrm{GSD}}_{i}  = J (s_{i-1}^{\alpha}+s_{i+1}^{\alpha})S^{\alpha} + D [(S^{z})^{2} - c_0 I].\label{eq:H_GSD_chain}
\end{equation}
This equation, and the definition $s_i^\alpha = \mathbf Z_i^\dagger S^\alpha \mathbf Z_i$, are sufficient to close the Schr\"odinger dynamics, Eq.~\eqref{eq:Schrodinger}. Note, in particular, that we need not explicitly select all SU(3) generators $T^{\alpha}$; the spin matrices $S^\alpha$ defined in Eq.~\eqref{eq:spin_1_matrices} are sufficient. We will use the normalization $|\mathbf Z_i|^2=1$ appropriate to coherent states.

The $c_0$ term in~\eqref{eq:H_GSD_chain} ensures $\mathrm{tr}\,\mathfrak H^{\mathrm{GSD}}_i = 0$, as expected for an element of $\mathfrak{su}(N)$. Under the Schr\"{o}dinger dynamics, this constant shift has the effect of rescaling $\mathbf Z_i$ by a physically irrelevant pure phase. Note, however, that this constant shift may meaningfully affect the result of numerical integration.

\subsection{Dynamics of a single spin}

As a first test, we will consider the dynamics of a single spin, setting $J=0$. Here, GSD is exactly faithful to the true quantum dynamics, whereas LLD is only approximate.

For simplicity, we take the initial condition to be a pure dipole,
\begin{equation}
    \mathbf s(t=0) =
    \left[\begin{array}{c}
s^x_0 \\
s^y_0 \\
s^z_0
\end{array}\right] =
\left[\begin{array}{c}
\sin(\theta) \\
0\\
\cos(\theta)
\end{array}\right]
\end{equation}
where $\theta$ denotes the polar angle relative to the $z$-axis.

The LLD result is simple precession about the $z$-axis,
\begin{equation}
\mathbf{s}^\mathrm{LLD}(t) =\left[\begin{array}{c}
s^x_0 \cos(\omega^\mathrm{LLD} t) \\
s^x_0 \sin(\omega^\mathrm{LLD} t)\\
s^z_0
\end{array}\right],\label{eq:single_ion_ll}
\end{equation}
where $\omega^\mathrm{LLD} = 2 D s^z_0$ is the angular frequency of precession.

\begin{figure}
    \centering
    \includegraphics[width=0.95\columnwidth]{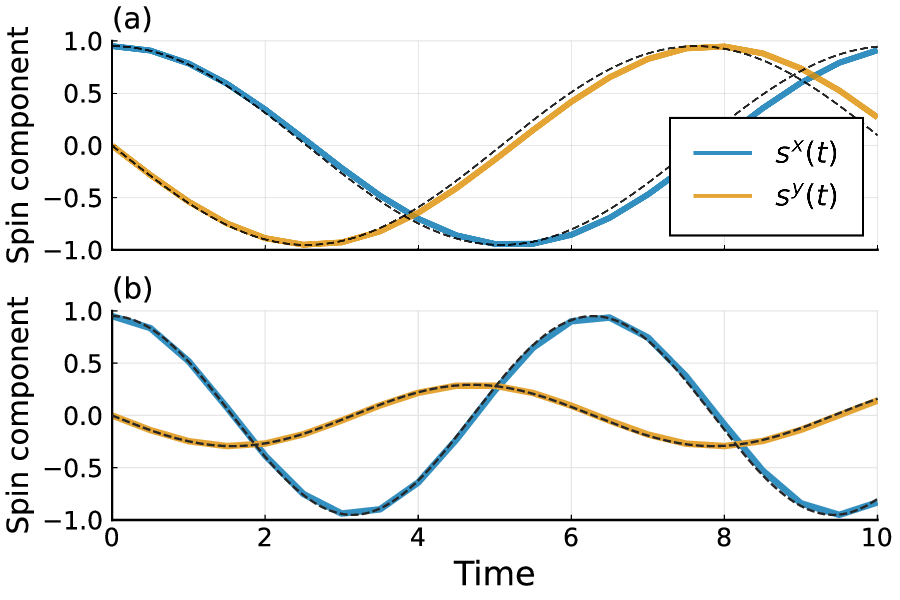}
    \caption{Numerical accuracy of integrated dipole trajectories for a single-spin Hamiltonian with an easy-axis anisotropy, $D=-1$. A large time step, $\Delta t = 0.5$, is selected to emphasize error. \textbf{(a)} LLD using the Schr\"odinger midpoint method in the spin-$\frac{1}{2}$ representation. \textbf{(b)} The quantum mechanically correct GSD, using the Schr\"odinger midpoint method applied to an $S=1$ spin (i.e., $N=3$). Dashed black lines indicate reference trajectories given in Eqs.~\eqref{eq:single_ion_ll} and Eq.~\eqref{eq:single_ion_gsd}. }
    \label{fig:fig1}
\end{figure}

In the single-site limit, the GSD describes the exact quantum evolution of expectation values. The trajectory for a single $S=1$ spin has been given analytically in Ref.~\onlinecite{Zhang21}. For our pure dipole initial condition, the dipole part of the trajectory is
\begin{equation}
\mathbf{s}^\mathrm{GSD}(t) =\left[\begin{array}{c}
s^x_0 \cos(\omega^\mathrm{GSD} t) \\
s^x_0 s^z_0 \sin(\omega^\mathrm{GSD} t)\\
s^z_0
\end{array}\right].\label{eq:single_ion_gsd}
\end{equation}
Like the dipole-only approximation, we again find precession, but here the frequency $\omega^\mathrm{GSD} = D$ is independent of the initial angle $\theta$. The $y$-component of the dipole is reduced by the factor $s^z_0$, such that the total spin dipole magnitude is no longer a constant of motion. This implies that weight oscillates between the three dipole and five quadrupole components of generalized spin $\mathbf n(t)$, and this effect is enhanced when the $z$-component of the spin is small. Recall that the quadratic Casimir $|\mathbf{n}(t)|^2$ is a constant of motion.

Figure~\ref{fig:fig1} illustrates the numerical integration of the LLD and GSD trajectories using the Schr\"odinger midpoint method for spin-$\frac{1}{2}$ and spin-1 respectively, and $\theta=2\pi/5$. A fairly large integration time-step $\Delta t = 0.5$ was selected to emphasize numerical error. For LLD, the spherical midpoint method~\cite{McLachlan17} is also applicable, and we observed its error to be about 2 times smaller (not shown). In all cases tested, energy conservation appears to be numerically exact under the Schr\"odinger midpoint integration scheme.

\subsection{Dynamics of a spin chain \label{subsec:benchmarks_LL}}

Next, we study a chain of $L=100$ spins with periodic boundary conditions, and a ferromagnetic Heisenberg interaction $J=-1$. We follow Ref.~\onlinecite{Viviani20} and select an initial state consisting of smoothly varying pure spin dipoles,
\begin{equation}
\mathbf{s}_{i}=\left[\begin{array}{c}
\cos(2\pi x_i^{2})\sin(2\pi x_i^{3})\\
\sin(2\pi x_i^{2})\sin(2\pi x_i^{3})\\
\cos(2\pi x_i^{3})
\end{array}\right] \label{eq:spin_initial}
\end{equation}
where $x_i=(i-1)/99$ for indices $i=1 \dots100$.

\subsubsection{Dynamics of a pure Heisenberg chain}

In the absence of the anisotropy term, $D = 0$, the classical Hamiltonian is purely a function of the spin dipole, and the LLD and GSD coincide.

\begin{figure}
    \centering
    \includegraphics[width=0.95\columnwidth]{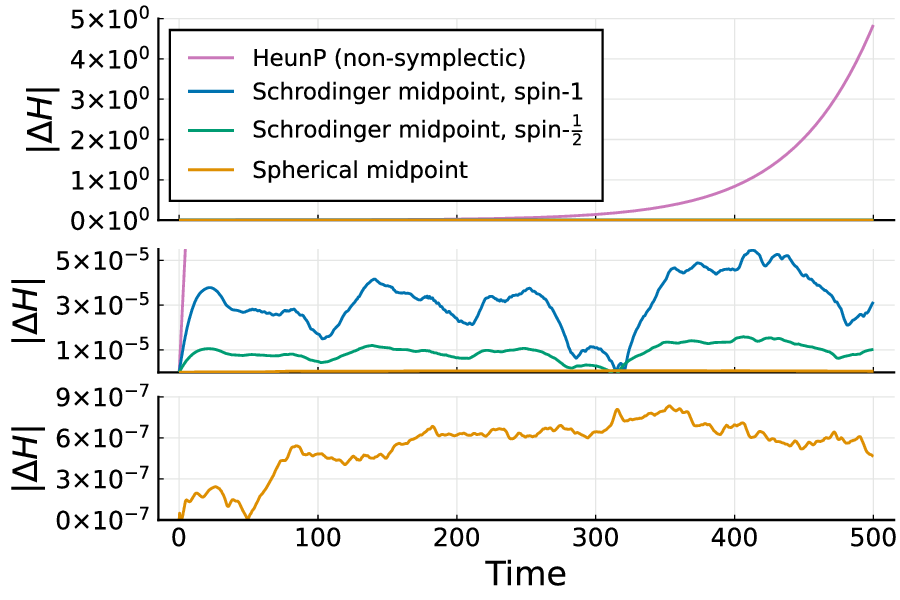}
    \caption{Energy fluctuations $\Delta H = H(t)-H(0)$ for the numerically integrated LLD of a pure Heisenberg spin chain, $J=-1$, absent anisotropy, $D=0$. We compare the following integration methods: the HeunP method applied to the Schr\"odinger equation, the Schr\"odinger midpoint method in the spin-$\frac{1}{2}$ and spin-1 representations, and the spherical midpoint method. The integration time-step is $\Delta t = 0.1$. The same data is plotted on three different energy scales to illustrate the huge variation between the integration schemes for this special model.}
    \label{fig:fig2}
\end{figure}

Figure~\ref{fig:fig2} illustrates the approximate conservation of energy for the various integration schemes applicable. Note that the Schr\"odinger midpoint method and the spherical midpoint method are both symplectic, and therefore prevent energy drift over large time-scales. In contrast, the HeunP scheme is non-symplectic, and exhibits a large, unphysical energy drift. Finally, we remark that the Schr\"odinger midpoint result with spin-1 precisely reproduces the curve shown in the top panel of the fourth figure in Ref.~\onlinecite{Viviani20}. To reproduce the bottom panel, we needed to integrate the spherical midpoint trajectory backwards in time.

\subsubsection{LLD including easy-axis anisotropy}

\begin{figure}
    \centering
    \includegraphics[width=0.95\columnwidth]{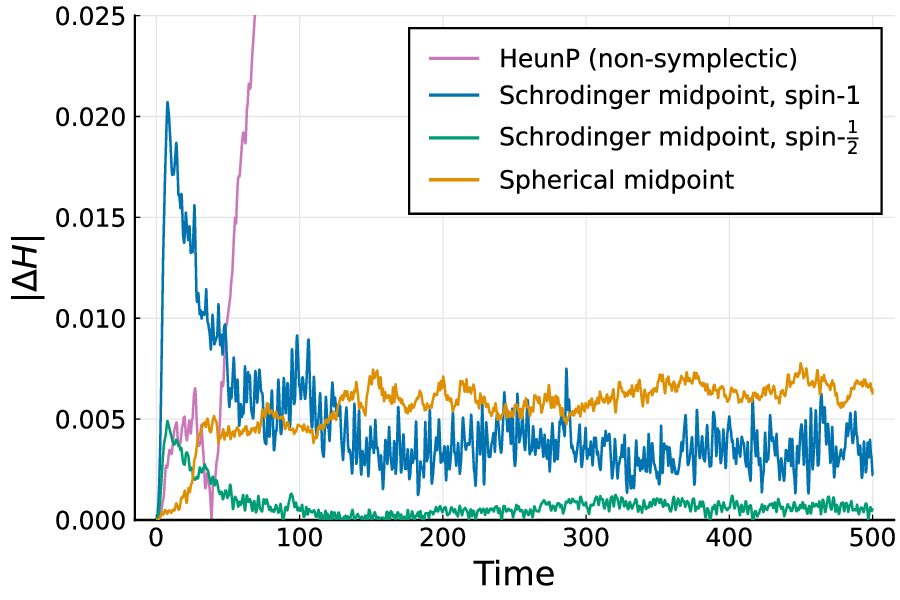}
    \caption{Energy fluctuations for the numerically integrated LLD of a Heisenberg spin chain with $J=-1$, now including an easy-axis anisotropy, $D=-1$. The numerical integration schemes are the same as in Fig.~\ref{fig:fig2}, but here we use a much smaller integration time-step, $\Delta t = 0.02$.}
    \label{fig:fig3}
\end{figure}

We now include an easy-axis anisotropy, $D=-1$, in the Heisenberg chain. With this additional term, the LLD and GSD classical limits deviate. First we will consider the LLD case.

Figure~\ref{fig:fig3} compares accuracy of four integration schemes. Relative to the $D=0$ case in Fig.~\ref{fig:fig2}, we observe much larger energy fluctuations, despite a significantly smaller time-step of $\Delta t = 0.02$ (down from $\Delta t = 0.1$). Energy fluctuations observed from the Schr\"odinger midpoint and spherical midpoint numerical integration schemes are now of the same order. In all cases tested, when applying the Schr\"odinger midpoint method to the LLD, the spin-$\frac{1}{2}$ representation is preferred over the spin-1 representation. There remains a tremendous advantage in using a symplectic integration scheme---the energy of the HeunP method continues to drift strongly.

\subsubsection{GSD including easy-axis anisotropy}

Our final numerical benchmark is the Heisenberg spin chain with easy-axis anisotropy $D=-1$ using the generalized spin dynamics. Here we must work with all three levels of the $S=1$ spins, which give rise to both dipole and quadrupole moments, and the traditional LLD numerical integration schemes do not apply.

\begin{figure}
    \centering
    \includegraphics[width=0.95\columnwidth]{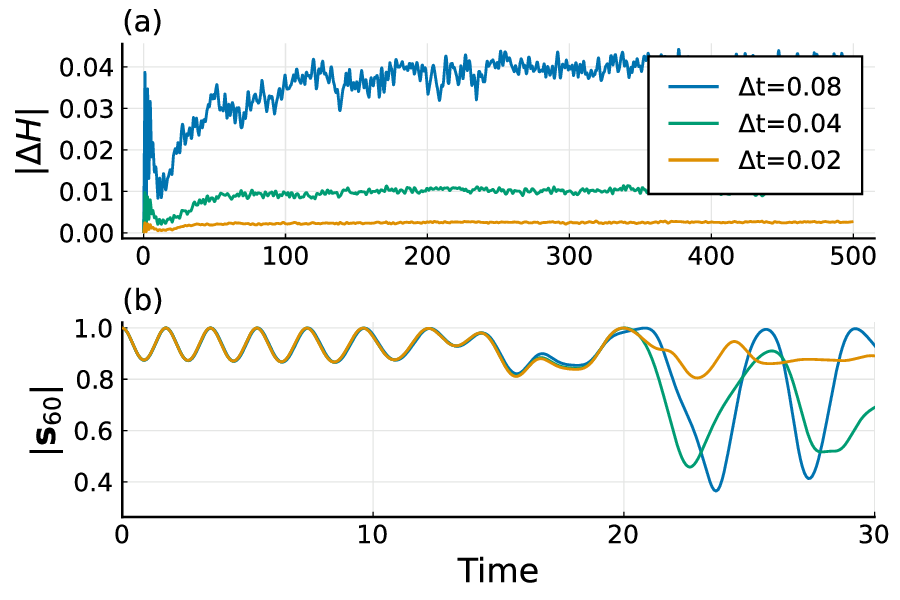}
    \caption{\textbf{(a)} Energy fluctuations for the numerically integrated GSD of a Heisenberg spin chain with $S=1$, $J=-1$ and easy-axis anisotropy, $D=-1$. We use the Schr\"odinger midpoint method with varying time-step $\Delta t$. \textbf{(b)} Time evolution of the dipole magnitude $|\mathbf s_i|$ for site index $i=60$.}
    \label{fig:fig4}
\end{figure}

Figure~\ref{fig:fig4} shows time integration using the Schr\"odinger midpoint method. The top panel illustrates that the energy fluctuations decrease approximately quadratically with time-step $\Delta t$, consistent with the second order accuracy of the implicit midpoint method~\cite{Hairer06}. The bottom panel shows the time evolution of the dipole magnitude $|\mathbf s_i|$ for site index $i=60$. Fluctuations in the dipole magnitude are possible because weight can be transferred to the quadrupole part of the generalized spins $\mathbf{n}_i$. All spins in the initial configuration, Eq.~\eqref{eq:spin_initial}, are pure dipoles, with a relatively slow variation along the spin chain. Therefore the initial dynamics is reasonably well approximated by the single spin limit, $J \approx 0$, previously considered in Fig.~\ref{fig:fig1}. At times $t \gtrsim 10$, however, high-frequency spatial variations in the spin chain propagate to site $i=60$. At this point, chaotic dynamics can be observed, and the three trajectories integrated using different $\Delta t$ quickly separate. As expected for a symplectic integrator, no significant energy drift is observed over arbitrarily long trajectory lengths. 

\section{Conclusions}

We have presented a numerical integration scheme, the Schr\"odinger midpoint method, that applies to the generalized spin dynamics, Eq.~\eqref{eq:gen_dynamics}. In the special case of the Landau-Lifshitz dynamics, Eq.~\eqref{eq:LL_dynamics}, this method reduces to previously known symplectic integrators~\cite{McLachlan15, Viviani20}. The Schr\"odinger midpoint method can be viewed as a special case of the Isospectral Midpoint Method~\cite{Modin20, Viviani20}, which applies to general Lie-Poisson systems. Compared to IMM, our approach is specialized to Eq.~\eqref{eq:gen_dynamics}, which arises as a classical limit of quantum mechanics, and has the numerical advantage of describing the evolution of a single eigenvector vector (the coherent spin state) rather than that of a full matrix. The method exactly respects the local symplectic structure of the Lie-Poisson system, or equivalently, the \emph{global} symplectic structure of the Schr\"odinger equation, which may be understood as a canonical Hamiltonian system. We anticipate that this method will be of broad applicability to the study of spin $S > \frac{1}{2}$ systems with strong single-ion anisotropy, for which the spin quadrupole (and perhaps higher-order) moments cannot be ignored~\cite{Zhang21, Bai21, Akagi21, Remund22, Amari22, Zhang22}.

\acknowledgments

We thank Martin Mourigal and Ying Wai Li for insightful
discussions. This work was supported by the U.S. Department of Energy, Office of Science, Office of Basic Energy Sciences, under Award No.~DE-SC0022311.

\section*{Code availability}

Code to reproduce all plots is available online at \url{https://github.com/SunnySuite/SchrodingerMidpoint.jl}. The methods we present are also implemented in \texttt{Sunny}, an open-source code for simulating general spin systems~\cite{sunny}.

\appendix

\section{Spin operators as representations of SU(2)\label{sec:spin_rep}}

Here we review some properties of the irreducible representations
of SU(2), which serve as quantum spin operators. The SU(2) irreps
are conventionally labeled by a spin index $S\in\{\frac{1}{2},1,\frac{3}{2},\dots\}$,
and have dimension $N=2S+1$.

For spin $S=1/2$, a conventional basis for the spin matrices is $S^\alpha = \sigma^\alpha / 2$, where $\sigma^\alpha$ are the Pauli matrices,

\begin{widetext}

\begin{equation}
S^{x}=\frac{1}{2}\left[\begin{array}{cc}
0 & 1\\
1 & 0
\end{array}\right],\,S^{y}=\frac{1}{2}\left[\begin{array}{cc}
0 & -\mathrm{i}\\
\mathrm{i} & 0
\end{array}\right],\,S^{z}=\frac{1}{2}\left[\begin{array}{cc}
1 & 0\\
0 & -1
\end{array}\right].
\end{equation}

More generally, it is convenient to work in a basis with $S^z$ diagonal. For spin $S=1$, 
\begin{equation}
S^{x}=\frac{1}{\sqrt{2}}\left[\begin{array}{ccc}
0 & 1 & 0\\
1 & 0 & 1\\
0 & 1 & 0
\end{array}\right],\,S^{y}=\frac{1}{\sqrt{2}}\left[\begin{array}{ccc}
0 & -\mathrm{i} & 0\\
\mathrm{i} & 0 & -\mathrm{i}\\
0 & \mathrm{i} & 0
\end{array}\right],\,S^{z}=\left[\begin{array}{ccc}
1 & 0 & 0\\
0 & 0 & 0\\
0 & 0 & -1
\end{array}\right]. \label{eq:spin_1_matrices}
\end{equation}
This pattern generalizes to arbitrary spin $S$,

\begin{equation}
S^{x}=\left[\begin{array}{cccc}
0 & a_{1}\\
a_{1} & 0 & \ddots\\
 & \ddots & \ddots & a_{N-1}\\
 &  & a_{N-1} & 0
\end{array}\right],\,S^{y}=\left[\begin{array}{cccc}
0 & -\im a_{1}\\
\im a_{1} & 0 & \ddots\\
 & \ddots & \ddots & -\im a_{N-1}\\
 &  & \im a_{N-1} & 0
\end{array}\right],\,S^{z}=\left[\begin{array}{cccc}
S\\
 & S-1\\
 &  & \ddots\\
 &  &  & -S
\end{array}\right],\label{eq:spin_rep_gen}
\end{equation}
where the off-diagonal elements, $a_{j}=\frac{1}{2}\sqrt{2(S+1)j-j(j+1)},$
satisfy the symmetry $a_{j}=a_{N-j}$. These coefficients also enter into raising and lowering operators, defined as $S_{\pm} = S^x \pm i S^y$.
\end{widetext}

The spin matrices $S^{\alpha}$ define the action of the quantum spin
operators. The many-body spin operator $\hat{S}_{i}^{\alpha}$ is
defined to act only on the $i$th local Hilbert space,
\[
\hat{S}_{i}^{\alpha}=\underset{\textrm{\ensuremath{i}th term}}{\underbrace{I\otimes\dots\otimes\hat{S}^{\alpha}}}\otimes\dots\otimes I.
\]
The local operator $\hat{S}^{\alpha}$ is defined by its action $\langle e_{a}|$$\hat{S}^{\alpha}|e_{b}\rangle=(S^{\alpha})_{ab}$
in some basis $|e_{1}\rangle,\dots,|e_{N}\rangle$. The operators
$\hat{S}_{i}^{\alpha}$ share all mathematical properties of the matrices
$S^{\alpha}$, to be listed below.

By construction, the spin matrices satisfy the commutation relations,
\begin{equation}
[S^{\alpha},S^{\beta}]=\im \epsilon_{\alpha\beta\gamma}S^{\gamma},\label{eq:su2_bracket_appendix}
\end{equation}
which gives rise to the same commutation relations, Eq.~(\ref{eq:su2_commut}),
for the quantum spin operators.

The $N$ eigenvalues for each spin matrix $S^{\alpha}$ run from $m=-S,\dots,S$.
The total angular momentum for each spin $i$ is scalar, i.e., 
\begin{equation}
|\mathbf{S}|^{2}=(S^{x})^{2}+(S^{y})^{2}+(S^{z})^{2}=S(S+1)I,
\end{equation}
with $I$ the identity matrix. This polynomial is the sole Casimir
of $\mathfrak{su}(2)$.

The spin matrices satisfy the orthonormality condition 
\begin{equation}
\mathrm{tr}\,S^{\alpha}S^{\beta}=\tau\delta_{\alpha\beta},
\end{equation}
which is a special case of Eq.~(\ref{eq:Lie_ortho}). The normalization
constant is the sum of squares of eigenvalues, 

\begin{equation}
\tau=\sum_{m=-S}^{S}m^{2}=\frac{2}{3}S\left(S+\frac{1}{2}\right)(S+1).\label{eq:tau_def}
\end{equation}

\section{Review of generalized spin dynamics\label{sec:qm_review}}

Following Ref.~\onlinecite{Zhang21}, we here review how Eq.~(\ref{eq:gen_dynamics})
emerges as a classical limit of a many-body quantum system.

Consider a many-body quantum spin Hamiltonian $\hat {\mathcal H}$, such as the spin chain model of Eq.~(\ref{eq:H_spin_chain}). This Hamiltonian may involve spin operators $\hat S^\alpha$ with spin representation $S$, corresponding to local Hilbert space dimension $N = 2S + 1$. The choice $S > 1/2$ would be appropriate to model, e.g., the effective spin angular momentum associated with a collection of spins that have aligned according to Hund's rules.

% \begin{equation}
% \hat{\mathcal{H}}=\frac{1}{2}\sum_{i\neq j}\sum_{\alpha,\beta}J_{ij}^{\alpha\beta}\hat{S}_{i}^{\alpha}\hat{S}_{j}^{\beta}-D\sum_{i}(\hat{S}_{i}^{z})^{2},\label{eq:spin_hamiltonian}
% \end{equation}
% which includes pairwise exchange interactions and an easy axis anisotropy.

% In a real magnetic material, each site $i$ may contain multiple valence-shell
% electrons. According to Hund's rules, the spins locally align at low
% temperatures. It is reasonable to model the quantum state on each
% site as a single effective spin. To do so, we will employ spin operators
% $\hat{S}_{i}^{\alpha}$ for an appropriate irreducible representation
% of SU(2), of dimension $N=2s+1$, where $s$ is the spin index. Properties
% of the SU(2) irreps are listed in Appendix~\ref{sec:spin_rep}.

For $N > 2$, the Hamiltonian may include local anisotropy terms, such as $(\hat{S}_{i}^{z})^{2}$. Such terms are \emph{not}
a linear combination of the $\hat{S}_{i}^{\alpha}$, i.e., do not
generate usual rotations of the local spin state. Importantly, however,
any local physical operator \emph{can} be decomposed as a linear combination
of the $N^{2}-1$ generators $\hat{T}_{i}^{\alpha}$ of SU($N$)
in the fundamental representation, plus a constant shift.

Using the completeness of the generators $\hat{T}_{i}^{\alpha}$
in each local Hilbert space, we can write a general Hamiltonian
as a polynomial expansion,
\begin{equation}
\hat{\mathcal{H}}=\sum_i J_{(i,\alpha)}^{(1)}\hat{T}_{i}^{\alpha}+ \frac{1}{2}\sum_{i,j}J_{(i,\alpha),(j,\beta)}^{(2)}\hat{T}_{i}^{\alpha}\hat{T}_{j}^{\beta}+\dots,\label{eq:H_def}
\end{equation}
up to an irrelevant constant shift. Recall that summation over repeated Greek indices is implied.

Any coefficient $J_{[\dots]}^{(n)}$ that couples a site with itself
(e.g., a single-ion anisotropy term) can
be effectively absorbed into lower order coefficients $J_{[\dots]}^{(n-1)}$
by the completeness of the generators $\hat{T}_{i}^{\alpha}$ for
each site $i$. Therefore, without loss of generality, we require
that the coefficients couple only distinct sites, e.g.,
\begin{align}
J_{(i,\alpha),(j,\beta)}^{(2)} & \propto(1-\delta_{ij}).\label{eq:J_off_diag}
\end{align}
Since $\hat{T}_{i}^{\alpha}$ and $\hat{T}_{j}^{\beta}$ commute for
$i\neq j$, we also have freedom to symmetrize the coefficients, e.g.,
\begin{equation}
J_{(i,\alpha),(j,\beta)}^{(2)}=J_{(j,\beta),(i,\alpha)}^{(2)}.\label{eq:J_sym}
\end{equation}

The Hamiltonian $\hat{\mathcal{H}}$ determines the evolution of a
general quantum state,

\begin{equation}
\frac{\mathd}{\mathd t}|\psi\rangle=e^{-\im t\hat{\mathcal{H}}}|\psi\rangle,\label{eq:Schrodinger_many_body}
\end{equation}
where we take $\hbar=1$. The time evolution of an arbitrary expectation
value $\langle\hat{A}\rangle=\langle\psi|\hat{A}|\psi\rangle$ follows,

\begin{equation}
\im\frac{\mathrm{d}}{\mathrm{d}t}\langle\hat{A}\rangle=\langle[\hat{A},\hat{\mathcal{H}}]\rangle.\label{eq:dA_dt}
\end{equation}

To take the classical limit, we will ignore quantum entanglement and
approximate the time-evolving many-body state $|\psi\rangle$ as a
tensor product of coherent states of a given algebra,
\begin{equation}
|\psi\rangle\approx|Z\rangle=\bigotimes_{i=1}^{L}|Z_{i}\rangle.\label{eq:coherent}
\end{equation}
In the remainder of this section, our purpose is to demonstrate the
following: \emph{This approximation yields a self-consistent
and closed dynamics on local expectation values}, namely, Eq.~(\ref{eq:gen_dynamics}).

Product states allow the factorization of expectation values over distinct
sites,
\begin{equation}
\langle Z|\hat{T}_{i}^{\alpha}\hat{T}_{j}^{\beta}|Z\rangle=n_{i}^{\alpha}n_{j}^{\beta}.
\end{equation}
It follows that, under the assumption of Eq.~\eqref{eq:coherent}, the expected
energy $H=\langle\hat{\mathcal{H}}\rangle$ is

\begin{equation}
H=\sum_{i} J_{(i,\alpha)}^{(1)}n_{i}^{\alpha}+\frac{1}{2}\sum_{i,j}J_{(i,\alpha),(j,\beta)}^{(2)}n_{i}^{\alpha}n_{j}^{\beta}+\dots\label{eq:E_def}
\end{equation}
In other words, substituting $\hat{T}_{i}^{\alpha}\rightarrow n_{i}^{\alpha}$
in the quantum Hamiltonian $\hat{\mathcal{H}}$ yields the classical
Hamiltonian $H$.

Inserting the general Hamiltonian of Eq.~(\ref{eq:H_def}) into the
dynamics Eq.~(\ref{eq:dA_dt}) for a local operator $\hat{A}=\hat{A}_{k}$
yields,
\begin{align}
\im\frac{\mathrm{d}}{\mathrm{d}t}\langle\hat{A}_{k}\rangle & =\sum_i J_{(i,\alpha)}^{(1)}\langle[\hat{A}_{k},\hat{T}_{i}^{\alpha}]\rangle\nonumber \\
 & \quad+\frac{1}{2} \sum_{i,j} J_{(i,\alpha),(j,\beta)}^{(2)}\langle[\hat{A}_{k},\hat{T}_{i}^{\alpha}\hat{T}_{j}^{\beta}]\rangle+\dots
\end{align}
For the first term, we use
\begin{equation}
\sum_i J_{(i,\alpha)}^{(1)}[\hat{A}_{k},\hat{T}_{i}^{\alpha}]=J_{(k,\alpha)}^{(1)}[\hat{A}_{k},\hat{T}_{k}^{\alpha}].
\end{equation}
For the second term, Eq.~(\ref{eq:J_off_diag}) ensures $i\neq j$,
and we require either $k=i$ or $k=j$. It follows,
\begin{align}
\frac{1}{2} \sum_{i,j} & J_{(i,\alpha),(j,\beta)}^{(2)}  [\hat{A}_{k},\hat{T}_{i}^{\alpha}\hat{T}_{j}^{\beta}]\nonumber \\
 & =\frac{1}{2}[\hat{A}_{k},\hat{T}_{k}^{\alpha}] \sum_j \left(J_{(k,\alpha),(j,\beta)}^{(2)}\hat{T}_{j}^{\beta}+J_{(j,\alpha),(k,\beta)}^{(2)}\hat{T}_{j}^{\alpha}\right)\nonumber \\
 & =[\hat{A}_{k},\hat{T}_{k}^{\alpha}] \sum_j J_{(k,\alpha),(j,\beta)}^{(2)}\hat{T}_{j}^{\beta}.
\end{align}
In the second step, we used the symmetrization convention of Eq.~(\ref{eq:J_sym}).
Combining results, and using again the approximation Eq.~(\ref{eq:coherent})
to factorize the expectation values on distinct sites, we find
\begin{equation}
\im\frac{\mathrm{d}}{\mathrm{d}t}\langle\hat{A}_{k}\rangle=\langle[\hat{A}_{k},\hat{T}_{k}^{\alpha}]\rangle\Big(J_{(k,\alpha)}^{(1)}+\sum_j J_{(k,\alpha),(j,\beta)}^{(2)}\langle\hat{T}_{j}^{\beta}\rangle+\dots\Big).
\end{equation}
The second factor on the right-hand side is $\partial H/\partial n_{k}^{\alpha}$,
and this result holds to all expansion orders. Relabeling $(k,\alpha)\rightarrow(i,\beta)$,
the result is
\begin{equation}
\im\frac{\mathrm{d}}{\mathrm{d}t}\langle\hat{A}_{i}\rangle=\langle[\hat{A}_{i},\hat{T}_{i}^{\beta}]\rangle\frac{\partial H}{\partial n_{i}^{\beta}},\label{eq:A_expect_heisenberg}
\end{equation}
which is valid for any local operator $\hat{A}_{i}$. Selecting $\hat{A}_{i}=\hat{T}_{i}^{\alpha}$
and using Eq.~(\ref{eq:bracket_op}), we reproduce the generalized
spin dynamics, Eq.~(\ref{eq:gen_dynamics}),
\begin{equation}
\frac{\mathrm{d}}{\mathrm{d}t}n_{i}^{\alpha}=f_{\alpha,\beta,\gamma}\frac{\partial H}{\partial n_{i}^{\beta}}n_{i}^{\gamma}.\label{eq:gen_dynamics_2}
\end{equation}

Although we motivated the form of $\hat{\mathcal{H}}$ in Eq.~(\ref{eq:H_def})
using the language of spin systems, this Hamiltonian is in fact fully
general. We could have taken each local Hilbert space $i$ to represent,
e.g., a tensor product space of \emph{multiple }quantum spins. The
time evolution of classical expectation values, Eq.~(\ref{eq:gen_dynamics_2}),
would then capture quantum entanglement effects within each local
Hilbert space.

\section{Hamiltonian structure of the Schr\"{o}dinger equation\label{sec:schro_canonical}}

Each $\mathfrak{su}(N)$ generator can be decomposed into its purely
real and imaginary parts,
\begin{equation}
T^{\alpha}=A^{\alpha}+\im B^{\alpha}.\label{eq:T_parts}
\end{equation}
Because $T^{\alpha}$ is Hermitian, it follows that $A^{\alpha}$
and $B^{\alpha}$ are symmetric and antisymmetric, respectively.

Substituting this decomposition into the Schr\"{o}dinger equation of Eq.~(\ref{eq:Schrodinger}),
we find
\begin{equation}
\frac{\mathd}{\mathd t}\mathbf{Z}_{i}=\frac{\partial H}{\partial n_{i}^{\alpha}}\left(-\im A^{\alpha}+B^{\alpha}\right)\mathbf{Z}_{i}.
\end{equation}
Decomposing the state vector into real and imaginary parts,

\begin{equation}
\mathbf{Z}_{i}=\frac{1}{\sqrt{2}}\left(\mathbf{p}_{i}-\im\mathbf{q}_{i}\right),
\end{equation}
allows formulation of the Schr\"{o}dinger equation as an entirely real
dynamics,

\begin{align}
\frac{\mathd \mathbf{p}_{i}}{\mathd t} & =\frac{\partial H}{\partial n_{i}^{\alpha}}\left(-A^{\alpha}\mathbf{q}_{i}+B^{\alpha}\mathbf{p}_{i}\right)\label{eq:schr_p}\\
\frac{\mathd \mathbf{q}_{i}}{\mathd t} & =\frac{\partial H}{\partial n_{i}^{\alpha}}\left(A^{\alpha}\mathbf{p}_{i}+B^{\alpha}\mathbf{q}_{i}\right).\label{eq:schr_q}
\end{align}

Expectation values $n_{i}^{\alpha}=\mathbf{Z}_{i}^{\dagger}T^{\alpha}\mathbf{Z}_{i}$
may be written,
\begin{align}
n_{i}^{\alpha} & =\frac{1}{2}\left(\mathbf{p}_{i}-\im\mathbf{q}_{i}\right)^{\dagger}T^{\alpha}\left(\mathbf{p}_{i}-\im\mathbf{q}_{i}\right).
\end{align}
Expanding the right-hand side, and decomposing $T^{\alpha}$ into
its symmetric and antisymmetric parts, we find
\begin{align}
n_{i}^{\alpha} & =\frac{1}{2}\mathbf{p}_{i}^{T}A^{\alpha}\mathbf{p}_{i}+\frac{1}{2}\mathbf{q}_{i}^{T}A^{\alpha}\mathbf{q}_{i}\nonumber \\
 & \quad\quad-\frac{1}{2}\mathbf{q}_{i}^{T}B^{\alpha}\mathbf{p}_{i}+\frac{1}{2}\mathbf{p}_{i}^{T}B^{\alpha}\mathbf{q}_{i}.
\end{align}
Differentiation yields
\begin{align}
\frac{\partial n_{i}^{\alpha}}{\partial\mathbf{p}_{i}} & =A^{\alpha}\mathbf{p}_{i}+B^{\alpha}\mathbf{q}_{i}\\
\frac{\partial n_{i}^{\alpha}}{\partial\mathbf{q}_{i}} & =A^{\alpha}\mathbf{q}_{i}-B^{\alpha}\mathbf{p}_{i}.
\end{align}
Energy derivatives can now be evaluated. Using the chain rule,
\begin{align}
\frac{\partial H}{\partial\mathbf{p}_{i}} & = \frac{\partial H}{\partial n_{i}^{\alpha}}\left(A^{\alpha}\mathbf{p}_{i}+B^{\alpha}\mathbf{q}_{i}\right)\\
\frac{\partial H}{\partial\mathbf{q}_{i}} & = \frac{\partial H}{\partial n_{i}^{\alpha}}\left(A^{\alpha}\mathbf{q}_{i}-B^{\alpha}\mathbf{p}_{i}\right).
\end{align}
Inserting these results into Eqs.~(\ref{eq:schr_p}) and~(\ref{eq:schr_q}),
we find that the Schr\"{o}dinger equation is described by Hamilton's equations
of motion,
\begin{equation}
\frac{\mathd\mathbf{p}_{i}}{\mathd t}=-\frac{\partial H}{\partial\mathbf{q}_{i}},\quad\frac{\mathd\mathbf{q}_{i}}{\mathd t}=+\frac{\partial H}{\partial\mathbf{p}_{i}}.
\end{equation}

\section{Derivation of the Schr\"{o}dinger midpoint formulas for LLD \label{sec:schr_ll_derivation}}

In this Appendix, we derive the results stated in Sec.~\ref{sec:schr_ll}. The LLD of spin dipoles, Eq.~\eqref{eq:LL_dynamics}, may be formulated as a Schr\"{o}dinger dynamics, Eq.~\eqref{eq:Schrodinger}, involving generators $S^\alpha$ of SU(2) in some representation.  The midpoint method applied to the Schr\"{o}dinger dynamics can be interpreted as two half time-steps, Eq.~\eqref{eq:Z_mid_back} and~\eqref{eq:Z_mid_forward}. Consider the former, from which we determine
\begin{equation}
s_{i}^{\alpha} =\tilde{\mathbf{Z}}_{i}^{\dagger}\left(S^{\alpha}+\im\frac{\Delta t}{2}[S^{\alpha},\tilde{\mathfrak{H}}_{i}]-\frac{\Delta t^{2}}{4}\tilde{\mathfrak{H}}_{i}S^{\alpha}\tilde{\mathfrak{H}}_{i}\right)\tilde{\mathbf{Z}}_{i},\label{eq:imm_half_update}
\end{equation}
where $s^\alpha_{i} = \mathbf{Z}_{i}^{\dagger}S^{\alpha}\mathbf{Z}_{i}$.

The first term is an expectation value, $\tilde{s}_i^{\alpha}=\tilde{\mathbf{Z}}_{i}^{\dagger}S^{\alpha}\tilde{\mathbf{Z}}_{i}$.
The second term may be expanded using Eqs.~\eqref{eq:frak_H} and~\eqref{eq:su2_bracket_appendix},
\begin{equation}
\im[S^{\alpha},\tilde{\mathfrak{H}}] = \im \frac{\partial H}{\partial\tilde{s}_{i}^{\beta}}[S^{\alpha},S^{\beta}] = - \epsilon_{\alpha \beta \gamma}  \frac{\partial H}{\partial\tilde{s}^\beta_i} S^\gamma.
\end{equation}
In vector notation,
\begin{equation}
\im \tilde{\mathbf Z}_i^\dagger  [\mathbf S,\tilde{\mathfrak{H}}] \tilde{\mathbf Z}_i =  \tilde{\mathbf{s}}_i \times \frac{\partial H}{\partial\tilde{\mathbf s}_i}.
\end{equation}

To evaluate the third term, we must incorporate more information about the
matrix representation of the generators $S^{\alpha}$. Let us introduce the notation,
\begin{equation}
    \tilde{\mathfrak H}_i =  \frac{\partial H}{\partial\tilde{s}^\beta_{i}} S^\beta = \mathbf a \cdot \mathbf S
\end{equation}

In the special
cases of the spin-$\frac{1}{2}$ and spin-1 representations, the matrix spin operators satisfy
\begin{equation}
(\mathbf{a}\cdot\mathbf{S}) \mathbf{S} (\mathbf{a}\cdot\mathbf{S})=\begin{cases}
\frac{\mathbf{a} (\mathbf{a}\cdot\mathbf{S})}{2}-\frac{|\mathbf{a}|^{2} \mathbf{S}}{4} & \textrm{(spin-$\frac{1}{2}$)}\\
\mathbf{a} (\mathbf{a}\cdot\mathbf{S}) & \textrm{(spin-1)}
\end{cases},
\end{equation}
valid for any $\mathbf{a} \in \mathbb{R}^3$.

It follows,
\begin{equation}
\tilde{\mathfrak{H}}\mathbf{S}\tilde{\mathfrak{H}}=\begin{cases}
\frac{1}{2}\frac{\partial H}{\partial\tilde{\mathbf{s}}_{i}}\left(\frac{\partial H}{\partial\tilde{\mathbf{s}}_{i}}\cdot\mathbf{S}\right)-\frac{1}{4}\left|\frac{\partial H}{\partial\tilde{\mathbf{s}}_{i}}\right|^{2}\mathbf{S} & \textrm{(spin-$\frac{1}{2}$)}\\
\frac{\partial H}{\partial\tilde{\mathbf{s}}_{i}}\left(\frac{\partial H}{\partial\tilde{\mathbf{s}}_{i}}\cdot\mathbf{S}\right) & \textrm{(spin-1)}
\end{cases}.
\end{equation}
Inserting these results into Eq.~(\ref{eq:imm_half_update}) yields Eq.~\eqref{eq:sch_ll_1}. A similar argument yields Eq.~\eqref{eq:sch_ll_2}. Combined, these equations provide a closed-form update rule entirely in terms of the spin dipoles.

\bibliographystyle{apsrev4-1}

\begin{thebibliography}{33}%
\makeatletter
\providecommand \@ifxundefined [1]{%
 \@ifx{#1\undefined}
}%
\providecommand \@ifnum [1]{%
 \ifnum #1\expandafter \@firstoftwo
 \else \expandafter \@secondoftwo
 \fi
}%
\providecommand \@ifx [1]{%
 \ifx #1\expandafter \@firstoftwo
 \else \expandafter \@secondoftwo
 \fi
}%
\providecommand \natexlab [1]{#1}%
\providecommand \enquote  [1]{``#1''}%
\providecommand \bibnamefont  [1]{#1}%
\providecommand \bibfnamefont [1]{#1}%
\providecommand \citenamefont [1]{#1}%
\providecommand \href@noop [0]{\@secondoftwo}%
\providecommand \href [0]{\begingroup \@sanitize@url \@href}%
\providecommand \@href[1]{\@@startlink{#1}\@@href}%
\providecommand \@@href[1]{\endgroup#1\@@endlink}%
\providecommand \@sanitize@url [0]{\catcode `\\12\catcode `\$12\catcode
  `\&12\catcode `\#12\catcode `\^12\catcode `\_12\catcode `\%12\relax}%
\providecommand \@@startlink[1]{}%
\providecommand \@@endlink[0]{}%
\providecommand \url  [0]{\begingroup\@sanitize@url \@url }%
\providecommand \@url [1]{\endgroup\@href {#1}{\urlprefix }}%
\providecommand \urlprefix  [0]{URL }%
\providecommand \Eprint [0]{\href }%
\providecommand \doibase [0]{http://dx.doi.org/}%
\providecommand \selectlanguage [0]{\@gobble}%
\providecommand \bibinfo  [0]{\@secondoftwo}%
\providecommand \bibfield  [0]{\@secondoftwo}%
\providecommand \translation [1]{[#1]}%
\providecommand \BibitemOpen [0]{}%
\providecommand \bibitemStop [0]{}%
\providecommand \bibitemNoStop [0]{.\EOS\space}%
\providecommand \EOS [0]{\spacefactor3000\relax}%
\providecommand \BibitemShut  [1]{\csname bibitem#1\endcsname}%
\let\auto@bib@innerbib\@empty
%</preamble>
\bibitem [{\citenamefont {Zhang}\ and\ \citenamefont
  {Batista}(2021)}]{Zhang21}%
  \BibitemOpen
  \bibfield  {author} {\bibinfo {author} {\bibfnamefont {H.}~\bibnamefont
  {Zhang}}\ and\ \bibinfo {author} {\bibfnamefont {C.~D.}\ \bibnamefont
  {Batista}},\ }\href {\doibase 10.1103/PhysRevB.104.104409} {\bibfield
  {journal} {\bibinfo  {journal} {Phys. Rev. B}\ }\textbf {\bibinfo {volume}
  {104}},\ \bibinfo {pages} {104409} (\bibinfo {year} {2021})}\BibitemShut
  {NoStop}%
\bibitem [{\citenamefont {Zapf}\ \emph {et~al.}(2006)\citenamefont {Zapf},
  \citenamefont {Zocco}, \citenamefont {Hansen}, \citenamefont {Jaime},
  \citenamefont {Harrison}, \citenamefont {Batista}, \citenamefont
  {Kenzelmann}, \citenamefont {Niedermayer}, \citenamefont {Lacerda},\ and\
  \citenamefont {Paduan-Filho}}]{Zapf06}%
  \BibitemOpen
  \bibfield  {author} {\bibinfo {author} {\bibfnamefont {V.~S.}\ \bibnamefont
  {Zapf}}, \bibinfo {author} {\bibfnamefont {D.}~\bibnamefont {Zocco}},
  \bibinfo {author} {\bibfnamefont {B.~R.}\ \bibnamefont {Hansen}}, \bibinfo
  {author} {\bibfnamefont {M.}~\bibnamefont {Jaime}}, \bibinfo {author}
  {\bibfnamefont {N.}~\bibnamefont {Harrison}}, \bibinfo {author}
  {\bibfnamefont {C.~D.}\ \bibnamefont {Batista}}, \bibinfo {author}
  {\bibfnamefont {M.}~\bibnamefont {Kenzelmann}}, \bibinfo {author}
  {\bibfnamefont {C.}~\bibnamefont {Niedermayer}}, \bibinfo {author}
  {\bibfnamefont {A.}~\bibnamefont {Lacerda}}, \ and\ \bibinfo {author}
  {\bibfnamefont {A.}~\bibnamefont {Paduan-Filho}},\ }\href {\doibase
  10.1103/PhysRevLett.96.077204} {\bibfield  {journal} {\bibinfo  {journal}
  {Phys. Rev. Lett.}\ }\textbf {\bibinfo {volume} {96}},\ \bibinfo {pages}
  {077204} (\bibinfo {year} {2006})}\BibitemShut {NoStop}%
\bibitem [{\citenamefont {Do}\ \emph {et~al.}(2021)\citenamefont {Do},
  \citenamefont {Zhang}, \citenamefont {Williams}, \citenamefont {Hong},
  \citenamefont {Garlea}, \citenamefont {Rodriguez-Rivera}, \citenamefont
  {Jang}, \citenamefont {Cheong}, \citenamefont {Park}, \citenamefont {Batista}
  \emph {et~al.}}]{Do2020}%
  \BibitemOpen
  \bibfield  {author} {\bibinfo {author} {\bibfnamefont {S.-H.}\ \bibnamefont
  {Do}}, \bibinfo {author} {\bibfnamefont {H.}~\bibnamefont {Zhang}}, \bibinfo
  {author} {\bibfnamefont {T.~J.}\ \bibnamefont {Williams}}, \bibinfo {author}
  {\bibfnamefont {T.}~\bibnamefont {Hong}}, \bibinfo {author} {\bibfnamefont
  {V.~O.}\ \bibnamefont {Garlea}}, \bibinfo {author} {\bibfnamefont
  {J.}~\bibnamefont {Rodriguez-Rivera}}, \bibinfo {author} {\bibfnamefont
  {T.-H.}\ \bibnamefont {Jang}}, \bibinfo {author} {\bibfnamefont {S.-W.}\
  \bibnamefont {Cheong}}, \bibinfo {author} {\bibfnamefont {J.-H.}\
  \bibnamefont {Park}}, \bibinfo {author} {\bibfnamefont {C.~D.}\ \bibnamefont
  {Batista}},  \emph {et~al.},\ }\href {\doibase 10.1038/s41467-021-25591-7}
  {\bibfield  {journal} {\bibinfo  {journal} {Nature communications}\ }\textbf
  {\bibinfo {volume} {12}},\ \bibinfo {pages} {1} (\bibinfo {year}
  {2021})}\BibitemShut {NoStop}%
\bibitem [{\citenamefont {Bai}\ \emph {et~al.}(2021)\citenamefont {Bai},
  \citenamefont {Zhang}, \citenamefont {Dun}, \citenamefont {Zhang},
  \citenamefont {Huang}, \citenamefont {Zhou}, \citenamefont {Stone},
  \citenamefont {Kolesnikov}, \citenamefont {Ye}, \citenamefont {Batista},\
  and\ \citenamefont {Mourigal}}]{Bai21}%
  \BibitemOpen
  \bibfield  {author} {\bibinfo {author} {\bibfnamefont {X.}~\bibnamefont
  {Bai}}, \bibinfo {author} {\bibfnamefont {S.-S.}\ \bibnamefont {Zhang}},
  \bibinfo {author} {\bibfnamefont {Z.}~\bibnamefont {Dun}}, \bibinfo {author}
  {\bibfnamefont {H.}~\bibnamefont {Zhang}}, \bibinfo {author} {\bibfnamefont
  {Q.}~\bibnamefont {Huang}}, \bibinfo {author} {\bibfnamefont
  {H.}~\bibnamefont {Zhou}}, \bibinfo {author} {\bibfnamefont {M.~B.}\
  \bibnamefont {Stone}}, \bibinfo {author} {\bibfnamefont {A.~I.}\ \bibnamefont
  {Kolesnikov}}, \bibinfo {author} {\bibfnamefont {F.}~\bibnamefont {Ye}},
  \bibinfo {author} {\bibfnamefont {C.~D.}\ \bibnamefont {Batista}}, \ and\
  \bibinfo {author} {\bibfnamefont {M.}~\bibnamefont {Mourigal}},\ }\href
  {\doibase 10.1038/s41567-020-01110-1} {\bibfield  {journal} {\bibinfo
  {journal} {Nat. Phys.}\ }\textbf {\bibinfo {volume} {17}},\ \bibinfo {pages}
  {467} (\bibinfo {year} {2021})}\BibitemShut {NoStop}%
\bibitem [{\citenamefont {Jaime}\ \emph {et~al.}(2004)\citenamefont {Jaime},
  \citenamefont {Correa}, \citenamefont {Harrison}, \citenamefont {Batista},
  \citenamefont {Kawashima}, \citenamefont {Kazuma}, \citenamefont {Jorge},
  \citenamefont {Stern}, \citenamefont {Heinmaa}, \citenamefont {Zvyagin},
  \citenamefont {Sasago},\ and\ \citenamefont {Uchinokura}}]{Jaime04}%
  \BibitemOpen
  \bibfield  {author} {\bibinfo {author} {\bibfnamefont {M.}~\bibnamefont
  {Jaime}}, \bibinfo {author} {\bibfnamefont {V.~F.}\ \bibnamefont {Correa}},
  \bibinfo {author} {\bibfnamefont {N.}~\bibnamefont {Harrison}}, \bibinfo
  {author} {\bibfnamefont {C.~D.}\ \bibnamefont {Batista}}, \bibinfo {author}
  {\bibfnamefont {N.}~\bibnamefont {Kawashima}}, \bibinfo {author}
  {\bibfnamefont {Y.}~\bibnamefont {Kazuma}}, \bibinfo {author} {\bibfnamefont
  {G.~A.}\ \bibnamefont {Jorge}}, \bibinfo {author} {\bibfnamefont
  {R.}~\bibnamefont {Stern}}, \bibinfo {author} {\bibfnamefont
  {I.}~\bibnamefont {Heinmaa}}, \bibinfo {author} {\bibfnamefont {S.~A.}\
  \bibnamefont {Zvyagin}}, \bibinfo {author} {\bibfnamefont {Y.}~\bibnamefont
  {Sasago}}, \ and\ \bibinfo {author} {\bibfnamefont {K.}~\bibnamefont
  {Uchinokura}},\ }\href {\doibase 10.1103/PhysRevLett.93.087203} {\bibfield
  {journal} {\bibinfo  {journal} {Phys. Rev. Lett.}\ }\textbf {\bibinfo
  {volume} {93}},\ \bibinfo {pages} {087203} (\bibinfo {year}
  {2004})}\BibitemShut {NoStop}%
\bibitem [{\citenamefont {Qiu}\ \emph {et~al.}(2005)\citenamefont {Qiu},
  \citenamefont {Broholm}, \citenamefont {Ishiwata}, \citenamefont {Azuma},
  \citenamefont {Takano}, \citenamefont {Bewley},\ and\ \citenamefont
  {Buyers}}]{Qiu05}%
  \BibitemOpen
  \bibfield  {author} {\bibinfo {author} {\bibfnamefont {Y.}~\bibnamefont
  {Qiu}}, \bibinfo {author} {\bibfnamefont {C.}~\bibnamefont {Broholm}},
  \bibinfo {author} {\bibfnamefont {S.}~\bibnamefont {Ishiwata}}, \bibinfo
  {author} {\bibfnamefont {M.}~\bibnamefont {Azuma}}, \bibinfo {author}
  {\bibfnamefont {M.}~\bibnamefont {Takano}}, \bibinfo {author} {\bibfnamefont
  {R.}~\bibnamefont {Bewley}}, \ and\ \bibinfo {author} {\bibfnamefont
  {W.~J.~L.}\ \bibnamefont {Buyers}},\ }\href {\doibase
  10.1103/PhysRevB.71.214439} {\bibfield  {journal} {\bibinfo  {journal} {Phys.
  Rev. B}\ }\textbf {\bibinfo {volume} {71}},\ \bibinfo {pages} {214439}
  (\bibinfo {year} {2005})}\BibitemShut {NoStop}%
\bibitem [{\citenamefont {Okamoto}\ \emph {et~al.}(2013)\citenamefont
  {Okamoto}, \citenamefont {Nilsen}, \citenamefont {Attfield},\ and\
  \citenamefont {Hiroi}}]{Okamoto13}%
  \BibitemOpen
  \bibfield  {author} {\bibinfo {author} {\bibfnamefont {Y.}~\bibnamefont
  {Okamoto}}, \bibinfo {author} {\bibfnamefont {G.~J.}\ \bibnamefont {Nilsen}},
  \bibinfo {author} {\bibfnamefont {J.~P.}\ \bibnamefont {Attfield}}, \ and\
  \bibinfo {author} {\bibfnamefont {Z.}~\bibnamefont {Hiroi}},\ }\href
  {\doibase 10.1103/PhysRevLett.110.097203} {\bibfield  {journal} {\bibinfo
  {journal} {Phys. Rev. Lett.}\ }\textbf {\bibinfo {volume} {110}},\ \bibinfo
  {pages} {097203} (\bibinfo {year} {2013})}\BibitemShut {NoStop}%
\bibitem [{\citenamefont {Marsden}\ and\ \citenamefont
  {Ratiu}(1999)}]{Marsden99}%
  \BibitemOpen
  \bibfield  {author} {\bibinfo {author} {\bibfnamefont {J.~E.}\ \bibnamefont
  {Marsden}}\ and\ \bibinfo {author} {\bibfnamefont {T.~S.}\ \bibnamefont
  {Ratiu}},\ }\href@noop {} {\emph {\bibinfo {title} {Introduction to
  {{Mechanics}} and {{Symmetry}}, {{A Basic Exposition}} of {{Classical
  Mechanical Systems}}}}},\ \bibinfo {series} {Texts in {{Applied
  Mathematic}}}, Vol.~\bibinfo {volume} {17}\ (\bibinfo  {publisher} {{Springer
  Berlin Heidelberg}},\ \bibinfo {year} {1999})\BibitemShut {NoStop}%
\bibitem [{\citenamefont {McLachlan}\ \emph {et~al.}(2015)\citenamefont
  {McLachlan}, \citenamefont {Modin},\ and\ \citenamefont
  {Verdier}}]{McLachlan15}%
  \BibitemOpen
  \bibfield  {author} {\bibinfo {author} {\bibfnamefont {R.~I.}\ \bibnamefont
  {McLachlan}}, \bibinfo {author} {\bibfnamefont {K.}~\bibnamefont {Modin}}, \
  and\ \bibinfo {author} {\bibfnamefont {O.}~\bibnamefont {Verdier}},\ }\href
  {\doibase 10.1093/imanum/dru013} {\bibfield  {journal} {\bibinfo  {journal}
  {IMA Journal of Numerical Analysis}\ }\textbf {\bibinfo {volume} {35}},\
  \bibinfo {pages} {546} (\bibinfo {year} {2015})}\BibitemShut {NoStop}%
\bibitem [{\citenamefont {Modin}\ and\ \citenamefont
  {Viviani}(2020)}]{Modin20}%
  \BibitemOpen
  \bibfield  {author} {\bibinfo {author} {\bibfnamefont {K.}~\bibnamefont
  {Modin}}\ and\ \bibinfo {author} {\bibfnamefont {M.}~\bibnamefont
  {Viviani}},\ }\href {\doibase 10.1007/s10208-019-09428-w} {\bibfield
  {journal} {\bibinfo  {journal} {Found Comput Math}\ }\textbf {\bibinfo
  {volume} {20}},\ \bibinfo {pages} {889} (\bibinfo {year} {2020})}\BibitemShut
  {NoStop}%
\bibitem [{\citenamefont {Viviani}(2020)}]{Viviani20}%
  \BibitemOpen
  \bibfield  {author} {\bibinfo {author} {\bibfnamefont {M.}~\bibnamefont
  {Viviani}},\ }\href {\doibase 10.1007/s10543-019-00792-1} {\bibfield
  {journal} {\bibinfo  {journal} {Bit Numer Math}\ }\textbf {\bibinfo {volume}
  {60}},\ \bibinfo {pages} {741} (\bibinfo {year} {2020})}\BibitemShut
  {NoStop}%
\bibitem [{\citenamefont {Eng{\o}}\ and\ \citenamefont
  {Faltinsen}(2001)}]{Engo01}%
  \BibitemOpen
  \bibfield  {author} {\bibinfo {author} {\bibfnamefont {K.}~\bibnamefont
  {Eng{\o}}}\ and\ \bibinfo {author} {\bibfnamefont {S.}~\bibnamefont
  {Faltinsen}},\ }\href {\doibase 10.1137/S0036142999364212} {\bibfield
  {journal} {\bibinfo  {journal} {SIAM J. Numer. Anal.}\ }\textbf {\bibinfo
  {volume} {39}},\ \bibinfo {pages} {128} (\bibinfo {year} {2001})}\BibitemShut
  {NoStop}%
\bibitem [{\citenamefont {Hairer}\ \emph {et~al.}(2006)\citenamefont {Hairer},
  \citenamefont {Lubich},\ and\ \citenamefont {Wanner}}]{Hairer06}%
  \BibitemOpen
  \bibfield  {author} {\bibinfo {author} {\bibfnamefont {E.}~\bibnamefont
  {Hairer}}, \bibinfo {author} {\bibfnamefont {C.}~\bibnamefont {Lubich}}, \
  and\ \bibinfo {author} {\bibfnamefont {G.}~\bibnamefont {Wanner}},\
  }\href@noop {} {\emph {\bibinfo {title} {Geometric {{Numerical Integration}}:
  {{Structure-Preserving Algorithms}} for {{Ordinary Differential
  Equations}}}}},\ \bibinfo {series} {Springer {{Series}} in {{Computational
  Mathematics}}}, Vol.~\bibinfo {volume} {31}\ (\bibinfo  {publisher}
  {{Springer Berlin Heidelberg}},\ \bibinfo {year} {2006})\BibitemShut
  {NoStop}%
\bibitem [{\citenamefont {Huberman}\ \emph {et~al.}(2008)\citenamefont
  {Huberman}, \citenamefont {Tennant}, \citenamefont {Cowley}, \citenamefont
  {Coldea},\ and\ \citenamefont {Frost}}]{Huberman08}%
  \BibitemOpen
  \bibfield  {author} {\bibinfo {author} {\bibfnamefont {T.}~\bibnamefont
  {Huberman}}, \bibinfo {author} {\bibfnamefont {D.~A.}\ \bibnamefont
  {Tennant}}, \bibinfo {author} {\bibfnamefont {R.~A.}\ \bibnamefont {Cowley}},
  \bibinfo {author} {\bibfnamefont {R.}~\bibnamefont {Coldea}}, \ and\ \bibinfo
  {author} {\bibfnamefont {C.~D.}\ \bibnamefont {Frost}},\ }\href {\doibase
  10.1088/1742-5468/2008/05/p05017} {\bibfield  {journal} {\bibinfo  {journal}
  {Journal of Statistical Mechanics: Theory and Experiment}\ }\textbf {\bibinfo
  {volume} {2008}},\ \bibinfo {pages} {P05017} (\bibinfo {year}
  {2008})}\BibitemShut {NoStop}%
\bibitem [{\citenamefont {Iserles}\ \emph {et~al.}(2000)\citenamefont
  {Iserles}, \citenamefont {{Munthe-Kaas}}, \citenamefont {N{\o}rsett},\ and\
  \citenamefont {Zanna}}]{Iserles00}%
  \BibitemOpen
  \bibfield  {author} {\bibinfo {author} {\bibfnamefont {A.}~\bibnamefont
  {Iserles}}, \bibinfo {author} {\bibfnamefont {H.~Z.}\ \bibnamefont
  {{Munthe-Kaas}}}, \bibinfo {author} {\bibfnamefont {S.~P.}\ \bibnamefont
  {N{\o}rsett}}, \ and\ \bibinfo {author} {\bibfnamefont {A.}~\bibnamefont
  {Zanna}},\ }\href {\doibase 10.1017/S0962492900002154} {\bibfield  {journal}
  {\bibinfo  {journal} {Acta Numerica}\ }\textbf {\bibinfo {volume} {9}},\
  \bibinfo {pages} {215} (\bibinfo {year} {2000})}\BibitemShut {NoStop}%
\bibitem [{\citenamefont {Iserles}\ and\ \citenamefont
  {Quispel}(2016)}]{Iserles16}%
  \BibitemOpen
  \bibfield  {author} {\bibinfo {author} {\bibfnamefont {A.}~\bibnamefont
  {Iserles}}\ and\ \bibinfo {author} {\bibfnamefont {G.~R.~W.}\ \bibnamefont
  {Quispel}},\ }\href@noop {} {\bibfield  {journal} {\bibinfo  {journal}
  {arXiv:1602.07755 [math]}\ } (\bibinfo {year} {2016})},\ \Eprint
  {http://arxiv.org/abs/1602.07755} {arXiv:1602.07755 [math]} \BibitemShut
  {NoStop}%
\bibitem [{\citenamefont {{Munthe-Kaas}}(1998)}]{Munthe-Kaas98}%
  \BibitemOpen
  \bibfield  {author} {\bibinfo {author} {\bibfnamefont {H.}~\bibnamefont
  {{Munthe-Kaas}}},\ }\href {\doibase 10.1007/BF02510919} {\bibfield  {journal}
  {\bibinfo  {journal} {Bit Numer Math}\ }\textbf {\bibinfo {volume} {38}},\
  \bibinfo {pages} {92} (\bibinfo {year} {1998})}\BibitemShut {NoStop}%
\bibitem [{\citenamefont {Zhong}\ and\ \citenamefont
  {Marsden}(1988)}]{Zhong88}%
  \BibitemOpen
  \bibfield  {author} {\bibinfo {author} {\bibfnamefont {G.}~\bibnamefont
  {Zhong}}\ and\ \bibinfo {author} {\bibfnamefont {J.~E.}\ \bibnamefont
  {Marsden}},\ }\href {\doibase 10.1016/0375-9601(88)90773-6} {\bibfield
  {journal} {\bibinfo  {journal} {Physics Letters A}\ }\textbf {\bibinfo
  {volume} {133}},\ \bibinfo {pages} {134} (\bibinfo {year}
  {1988})}\BibitemShut {NoStop}%
\bibitem [{\citenamefont {Channell}\ and\ \citenamefont
  {Scovel}(1991)}]{Channell91}%
  \BibitemOpen
  \bibfield  {author} {\bibinfo {author} {\bibfnamefont {P.~J.}\ \bibnamefont
  {Channell}}\ and\ \bibinfo {author} {\bibfnamefont {J.~C.}\ \bibnamefont
  {Scovel}},\ }\href {\doibase 10.1016/0167-2789(91)90081-J} {\bibfield
  {journal} {\bibinfo  {journal} {Physica D: Nonlinear Phenomena}\ }\textbf
  {\bibinfo {volume} {50}},\ \bibinfo {pages} {80} (\bibinfo {year}
  {1991})}\BibitemShut {NoStop}%
\bibitem [{\citenamefont {McLachlan}(1993)}]{McLachlan93}%
  \BibitemOpen
  \bibfield  {author} {\bibinfo {author} {\bibfnamefont {R.~I.}\ \bibnamefont
  {McLachlan}},\ }\href {\doibase 10.1103/PhysRevLett.71.3043} {\bibfield
  {journal} {\bibinfo  {journal} {Phys. Rev. Lett.}\ }\textbf {\bibinfo
  {volume} {71}},\ \bibinfo {pages} {3043} (\bibinfo {year}
  {1993})}\BibitemShut {NoStop}%
\bibitem [{\citenamefont {McLachlan}\ and\ \citenamefont
  {Scovel}(1995)}]{McLachlan95}%
  \BibitemOpen
  \bibfield  {author} {\bibinfo {author} {\bibfnamefont {R.~I.}\ \bibnamefont
  {McLachlan}}\ and\ \bibinfo {author} {\bibfnamefont {C.}~\bibnamefont
  {Scovel}},\ }\href {\doibase 10.1007/BF01212956} {\bibfield  {journal}
  {\bibinfo  {journal} {J Nonlinear Sci}\ }\textbf {\bibinfo {volume} {5}},\
  \bibinfo {pages} {233} (\bibinfo {year} {1995})}\BibitemShut {NoStop}%
\bibitem [{\citenamefont {Krech}\ \emph {et~al.}(1998)\citenamefont {Krech},
  \citenamefont {Bunker},\ and\ \citenamefont {Landau}}]{Krech98}%
  \BibitemOpen
  \bibfield  {author} {\bibinfo {author} {\bibfnamefont {M.}~\bibnamefont
  {Krech}}, \bibinfo {author} {\bibfnamefont {A.}~\bibnamefont {Bunker}}, \
  and\ \bibinfo {author} {\bibfnamefont {D.~P.}\ \bibnamefont {Landau}},\
  }\href {\doibase 10.1016/S0010-4655(98)00009-5} {\bibfield  {journal}
  {\bibinfo  {journal} {Computer Physics Communications}\ }\textbf {\bibinfo
  {volume} {111}},\ \bibinfo {pages} {1} (\bibinfo {year} {1998})}\BibitemShut
  {NoStop}%
\bibitem [{\citenamefont {Omelyan}\ \emph {et~al.}(2001)\citenamefont
  {Omelyan}, \citenamefont {Mryglod},\ and\ \citenamefont {Folk}}]{Omelyan01}%
  \BibitemOpen
  \bibfield  {author} {\bibinfo {author} {\bibfnamefont {I.~P.}\ \bibnamefont
  {Omelyan}}, \bibinfo {author} {\bibfnamefont {I.~M.}\ \bibnamefont
  {Mryglod}}, \ and\ \bibinfo {author} {\bibfnamefont {R.}~\bibnamefont
  {Folk}},\ }\href {\doibase 10.1103/PhysRevLett.86.898} {\bibfield  {journal}
  {\bibinfo  {journal} {Phys. Rev. Lett.}\ }\textbf {\bibinfo {volume} {86}},\
  \bibinfo {pages} {898} (\bibinfo {year} {2001})}\BibitemShut {NoStop}%
\bibitem [{\citenamefont {Tranchida}\ \emph {et~al.}(2018)\citenamefont
  {Tranchida}, \citenamefont {Plimpton}, \citenamefont {Thibaudeau},\ and\
  \citenamefont {Thompson}}]{Tranchida18}%
  \BibitemOpen
  \bibfield  {author} {\bibinfo {author} {\bibfnamefont {J.}~\bibnamefont
  {Tranchida}}, \bibinfo {author} {\bibfnamefont {S.~J.}\ \bibnamefont
  {Plimpton}}, \bibinfo {author} {\bibfnamefont {P.}~\bibnamefont
  {Thibaudeau}}, \ and\ \bibinfo {author} {\bibfnamefont {A.~P.}\ \bibnamefont
  {Thompson}},\ }\href {\doibase 10.1016/j.jcp.2018.06.042} {\bibfield
  {journal} {\bibinfo  {journal} {Journal of Computational Physics}\ }\textbf
  {\bibinfo {volume} {372}},\ \bibinfo {pages} {406} (\bibinfo {year}
  {2018})}\BibitemShut {NoStop}%
\bibitem [{\citenamefont {McLachlan}\ \emph {et~al.}(2014)\citenamefont
  {McLachlan}, \citenamefont {Modin},\ and\ \citenamefont
  {Verdier}}]{McLachlan14}%
  \BibitemOpen
  \bibfield  {author} {\bibinfo {author} {\bibfnamefont {R.~I.}\ \bibnamefont
  {McLachlan}}, \bibinfo {author} {\bibfnamefont {K.}~\bibnamefont {Modin}}, \
  and\ \bibinfo {author} {\bibfnamefont {O.}~\bibnamefont {Verdier}},\ }\href
  {\doibase 10.1103/PhysRevE.89.061301} {\bibfield  {journal} {\bibinfo
  {journal} {Phys. Rev. E}\ }\textbf {\bibinfo {volume} {89}},\ \bibinfo
  {pages} {061301} (\bibinfo {year} {2014})}\BibitemShut {NoStop}%
\bibitem [{\citenamefont {McLachlan}\ \emph {et~al.}(2017)\citenamefont
  {McLachlan}, \citenamefont {Modin},\ and\ \citenamefont
  {Verdier}}]{McLachlan17}%
  \BibitemOpen
  \bibfield  {author} {\bibinfo {author} {\bibfnamefont {R.}~\bibnamefont
  {McLachlan}}, \bibinfo {author} {\bibfnamefont {K.}~\bibnamefont {Modin}}, \
  and\ \bibinfo {author} {\bibfnamefont {O.}~\bibnamefont {Verdier}},\ }\href
  {\doibase 10.1090/mcom/3153} {\bibfield  {journal} {\bibinfo  {journal}
  {Math. Comp.}\ }\textbf {\bibinfo {volume} {86}},\ \bibinfo {pages} {2325}
  (\bibinfo {year} {2017})}\BibitemShut {NoStop}%
\bibitem [{\citenamefont {Skubic}\ \emph {et~al.}(2008)\citenamefont {Skubic},
  \citenamefont {Hellsvik}, \citenamefont {Nordstr{\"o}m},\ and\ \citenamefont
  {Eriksson}}]{Skubic08}%
  \BibitemOpen
  \bibfield  {author} {\bibinfo {author} {\bibfnamefont {B.}~\bibnamefont
  {Skubic}}, \bibinfo {author} {\bibfnamefont {J.}~\bibnamefont {Hellsvik}},
  \bibinfo {author} {\bibfnamefont {L.}~\bibnamefont {Nordstr{\"o}m}}, \ and\
  \bibinfo {author} {\bibfnamefont {O.}~\bibnamefont {Eriksson}},\ }\href
  {\doibase 10.1088/0953-8984/20/31/315203} {\bibfield  {journal} {\bibinfo
  {journal} {J. Phys.: Condens. Matter}\ }\textbf {\bibinfo {volume} {20}},\
  \bibinfo {pages} {315203} (\bibinfo {year} {2008})}\BibitemShut {NoStop}%
\bibitem [{\citenamefont {Mentink}\ \emph {et~al.}(2010)\citenamefont
  {Mentink}, \citenamefont {Tretyakov}, \citenamefont {Fasolino}, \citenamefont
  {Katsnelson},\ and\ \citenamefont {Rasing}}]{Mentink10}%
  \BibitemOpen
  \bibfield  {author} {\bibinfo {author} {\bibfnamefont {J.~H.}\ \bibnamefont
  {Mentink}}, \bibinfo {author} {\bibfnamefont {M.~V.}\ \bibnamefont
  {Tretyakov}}, \bibinfo {author} {\bibfnamefont {A.}~\bibnamefont {Fasolino}},
  \bibinfo {author} {\bibfnamefont {M.~I.}\ \bibnamefont {Katsnelson}}, \ and\
  \bibinfo {author} {\bibfnamefont {T.}~\bibnamefont {Rasing}},\ }\href
  {\doibase 10.1088/0953-8984/22/17/176001} {\bibfield  {journal} {\bibinfo
  {journal} {J. Phys.: Condens. Matter}\ }\textbf {\bibinfo {volume} {22}},\
  \bibinfo {pages} {176001} (\bibinfo {year} {2010})}\BibitemShut {NoStop}%
\bibitem [{\citenamefont {Akagi}\ \emph {et~al.}(2021)\citenamefont {Akagi},
  \citenamefont {Amari}, \citenamefont {Sawado},\ and\ \citenamefont
  {Shnir}}]{Akagi21}%
  \BibitemOpen
  \bibfield  {author} {\bibinfo {author} {\bibfnamefont {Y.}~\bibnamefont
  {Akagi}}, \bibinfo {author} {\bibfnamefont {Y.}~\bibnamefont {Amari}},
  \bibinfo {author} {\bibfnamefont {N.}~\bibnamefont {Sawado}}, \ and\ \bibinfo
  {author} {\bibfnamefont {Y.}~\bibnamefont {Shnir}},\ }\href {\doibase
  10.1103/PhysRevD.103.065008} {\bibfield  {journal} {\bibinfo  {journal}
  {Phys. Rev. D}\ }\textbf {\bibinfo {volume} {103}},\ \bibinfo {pages}
  {065008} (\bibinfo {year} {2021})},\ \Eprint
  {http://arxiv.org/abs/2101.10566} {arXiv:2101.10566} \BibitemShut {NoStop}%
\bibitem [{\citenamefont {Remund}\ \emph {et~al.}(2022)\citenamefont {Remund},
  \citenamefont {Pohle}, \citenamefont {Akagi}, \citenamefont {Romh{\'a}nyi},\
  and\ \citenamefont {Shannon}}]{Remund22}%
  \BibitemOpen
  \bibfield  {author} {\bibinfo {author} {\bibfnamefont {K.}~\bibnamefont
  {Remund}}, \bibinfo {author} {\bibfnamefont {R.}~\bibnamefont {Pohle}},
  \bibinfo {author} {\bibfnamefont {Y.}~\bibnamefont {Akagi}}, \bibinfo
  {author} {\bibfnamefont {J.}~\bibnamefont {Romh{\'a}nyi}}, \ and\ \bibinfo
  {author} {\bibfnamefont {N.}~\bibnamefont {Shannon}},\ }\href@noop {}
  {\bibfield  {journal} {\bibinfo  {journal} {arXiv:2203.09819 [cond-mat]}\ }
  (\bibinfo {year} {2022})},\ \Eprint {http://arxiv.org/abs/2203.09819}
  {arXiv:2203.09819 [cond-mat]} \BibitemShut {NoStop}%
\bibitem [{\citenamefont {Amari}\ \emph {et~al.}(2022)\citenamefont {Amari},
  \citenamefont {Akagi}, \citenamefont {Gudnason}, \citenamefont {Nitta},\ and\
  \citenamefont {Shnir}}]{Amari22}%
  \BibitemOpen
  \bibfield  {author} {\bibinfo {author} {\bibfnamefont {Y.}~\bibnamefont
  {Amari}}, \bibinfo {author} {\bibfnamefont {Y.}~\bibnamefont {Akagi}},
  \bibinfo {author} {\bibfnamefont {S.~B.}\ \bibnamefont {Gudnason}}, \bibinfo
  {author} {\bibfnamefont {M.}~\bibnamefont {Nitta}}, \ and\ \bibinfo {author}
  {\bibfnamefont {Y.}~\bibnamefont {Shnir}},\ }\href@noop {} {\bibfield
  {journal} {\bibinfo  {journal} {arXiv:2204.01476 [cond-mat, physics:hep-th]}\
  } (\bibinfo {year} {2022})},\ \Eprint {http://arxiv.org/abs/2204.01476}
  {arXiv:2204.01476 [cond-mat, physics:hep-th]} \BibitemShut {NoStop}%
\bibitem [{\citenamefont {Zhang}\ \emph {et~al.}(2022)\citenamefont {Zhang},
  \citenamefont {Wang}, \citenamefont {Dahlbom}, \citenamefont {Barros},\ and\
  \citenamefont {Batista}}]{Zhang22}%
  \BibitemOpen
  \bibfield  {author} {\bibinfo {author} {\bibfnamefont {H.}~\bibnamefont
  {Zhang}}, \bibinfo {author} {\bibfnamefont {Z.}~\bibnamefont {Wang}},
  \bibinfo {author} {\bibfnamefont {D.}~\bibnamefont {Dahlbom}}, \bibinfo
  {author} {\bibfnamefont {K.}~\bibnamefont {Barros}}, \ and\ \bibinfo {author}
  {\bibfnamefont {C.~D.}\ \bibnamefont {Batista}},\ }\href@noop {} {\bibfield
  {journal} {\bibinfo  {journal} {arXiv:2203.15248 [cond-mat]}\ } (\bibinfo
  {year} {2022})},\ \Eprint {http://arxiv.org/abs/2203.15248} {arXiv:2203.15248
  [cond-mat]} \BibitemShut {NoStop}%
\bibitem [{sun(2022)}]{sunny}%
  \BibitemOpen
  \href@noop {} {} (\bibinfo {year} {2022}),\ \bibinfo {note}
  {\url{https://github.com/SunnySuite/Sunny.jl/}}\BibitemShut {NoStop}%
\end{thebibliography}

\end{document}